\documentclass[pre,twocolumn,showkeys,showpacs,superscriptaddress]{revtex4-1}
\usepackage{graphicx,latexsym,amssymb,amsmath,amsbsy,color}

\begin{document}
\title{Search for Hyperuniformity in Mechanically Stable Packings of Frictionless Disks Above Jamming}
\author         {Yegang Wu}
\affiliation    {Department of Physics and Astronomy, University of Rochester, Rochester, New York 14627, USA}
\author{Peter Olsson}
\affiliation{Department of Physics, Ume{\aa} University, 901 87 Ume{\aa}, Sweden}
\author{S. Teitel}
\affiliation    {Department of Physics and Astronomy, University of Rochester, Rochester, New York 14627, USA}
\date{\today}

\begin{abstract}
We numerically simulate mechanically stable packings of soft-core, frictionless, bidisperse disks in two dimensions, {\em above} the jamming packing fraction $\phi_J$.  For configurations with a fixed isotropic global stress tensor, we investigate the fluctuations of the local packing fraction $\phi(\mathbf{r})$ to test whether such configurations display the hyperuniformity that has been claimed to exist exactly at $\phi_J$.  For our configurations, generated by a rapid quench protocol, we find that hyperuniformity persists only out to a finite length scale, and that this length scale appears to remain finite as the system stress decreases towards zero, i.e. towards the jamming transition.  Our result suggests that the presence of hyperuniformity at jamming may be sensitive to the specific protocol used to construct the jammed configurations.
\end{abstract}
\maketitle


\maketitle

\section{Introduction}

When a system of athermal ($T=0$) particles with only contact interactions is compressed, it seizes up into a rigid disordered solid at a critical value of the packing fraction $\phi_J$ known as the jamming transition \cite{Liu+Nagel,OHern,vanHecke}.  For a system of monodisperse frictionless spheres at $\phi_J$, it was observed numerically \cite{Donev1,Silbert} that density fluctuations appear to be suppressed on long length scales, with a structure function $S(\mathbf{q})$ (density-density correlation) that vanishes as $S(\mathbf{q})\sim |\mathbf{q}|$ when the wavevector $|\mathbf{q}|\to 0$.  This is in contrast to behavior in a normal liquid where $S(\mathbf{q}\to 0)\to$ constant. Such a system with suppressed density fluctuations has been denoted as ``hyperuniform" \cite{Torquato}.

However when spheres that were bidisperse or polydisperse in size where studied, this characteristic feature of $S(\mathbf{q})$ was no longer observed, and $S(\mathbf{q}\to 0)$ was found to be finite \cite{Xu,Kurita}.  It was then argued by Berthier et al. \cite{Berthier}, and by Zachary et al. \cite{Zachary2,Zachary3}, that in such size-disperse systems it is the fluctuations of the packing fraction $\phi$, rather than fluctuations of particle density, that are suppressed at $\phi_J$ (for monodisperse systems, packing fraction fluctuations and density fluctuations become the same at long wavelengths).  The presence of such hyperuniformity of the packing fraction at jamming would be important, as it would provide a purely structural means for distinguishing particles in a disordered jammed configuration from those in a liquid,
and perhaps provide a way to determine a diverging length scale as the jamming transition is approached \cite{Hopkins}.

In this work we consider mechanically stable packings of bidisperse, soft-core, frictionless, disks in two dimensions  at finite {\em isotropic} global stress {\em above} the jamming transition $\phi_J$.  Our configurations are generated by a rapid quench protocol.  We test these configurations for hyperuniformity using both real-space and wavevector-space methods.  
We find that hyperuniformity persists only out to a finite length scale, and that this length scale appears to remain finite as the system stress decreases towards zero, i.e. as one approaches the jamming transition. Moreover, we argue that measuring fluctuations at a given wavevector $q$ gives a better test of hyperuniformity than measuring fluctuations over a real-space window of length $R$, as the latter can be strongly effected by the fluctuations on all length scales smaller than $R$, whereas the former measures fluctuations specifically on the length scale $2\pi/q$.

The remainder of this paper is organized as follows.  In Sec.~\ref{secII} we define what we mean by the local packing fraction $\phi(\mathbf{r})$ and discuss the wavevector-dependent and real-space measures we will use to test for hyperuniformity.  In Sec.~\ref{SecModel} we describe the details of our numerical model and the minimization method we use to construct mechanically stable configurations at fixed isotropic global stress.  In Sec.~\ref{secResults} we present our numerical results.  In Sec.~\ref{secDiscuss} we discuss our results and make comparisons with recent works on this topic.  The Appendix provides further details about the accuracy of our numerical minimization method for constructing our configurations.

\section{Local packing fraction}
\label{secII}

In this section we define the quantities we will compute in order to test for hyperuniformity.  Here we define quantities as appropriate to a system of two dimensional circular disks, so as to match our numerical simulations, however the generalization to higher dimension or other shaped particles is straightforward.

Consider a polydisperse collection of $N$ disks in a system of total volume $V$, satisfying Lees-Edwards boundary conditions \cite{LeesEdwards}.  Disk $i$ has its center located at position $\mathbf{r}_i$ and has volume $v_i$ (in our two dimensional system we will use ``volume" to mean area).  The local particle density can then be written as,
\begin{equation}
n(\mathbf{r})=\sum_i\delta(\mathbf{r}-\mathbf{r}_i).
\end{equation}
Defining the Fourier transform,
\begin{equation}
n_\mathbf{q}=\int_Vd^2r\,\mathrm{e}^{i\mathbf{q}\cdot\mathbf{r}}n(\mathbf{r}),
\end{equation}
the structure function (density-density correlation) is,
\begin{equation}
S(\mathbf{q})\equiv\frac{1}{N}\langle n_\mathbf{q}n_{-\mathbf{q}}\rangle,
\end{equation}
where here, and henceforth, $\langle\dots\rangle$ denotes an average over independently quenched configurations.  For the bidisperse systems we study here, we expect $S(\mathbf{q})$ to approach a constant as $|\mathbf{q}|\to 0$ \cite{Xu,Kurita,Berthier,Zachary2, Zachary3}.

The {\em global} packing fraction of the system is defined as,
\begin{equation}
\phi\equiv\frac{1}{V}\sum _i v_i.
\label{phig}
\end{equation}
For the {\em local} packing fraction $\phi(\mathbf{r})$, two slightly different definitions have been proposed in the literature.  Zachary et al. \cite{Zachary2,Zachary3} use a definition that is equivalent to,
\begin{equation}
\mathrm{definition\,I:}\qquad\phi(\mathbf{r})=\sum_i\Delta_i(\mathbf{r}-\mathbf{r}_i),
\label{edefI}
\end{equation}
where the indicator function $\Delta_i(\mathbf{r})$ is such that for a particle centered at the origin,
\begin{equation}
\Delta_i(\mathbf{r})=\left\{
\begin{array}{ll}
1,&\mathrm{if\,}\mathbf{r}\,\mathrm{lies\,within\,the\,area\,of\,the\,particle}\\
0,&\mathrm{otherwise}
\end{array}
\right.
\end{equation}
so that $\int_V d^2r\,\Delta_i(\mathbf{r}_i)=v_i$.

Berthier et al. \cite{Berthier} use a definition \cite{note} that is equivalent to,
\begin{equation}
\mathrm{definition\,II:}\qquad\phi(\mathbf{r})=\sum_i v_i\delta(\mathbf{r}-\mathbf{r}_i).
\label{edefII}
\end{equation}
Both definitions give correctly the global packing fraction of Eq.~(\ref{phig}), 
\begin{equation}
\frac{1}{V}\int_Vd^2r\,\phi(\mathbf{r})=\frac{1}{V}\sum_i v_i=\phi.
\end{equation}
Definition I spreads the weight of each particle uniformly over its area, while definition II treats each particle as a point object with weight equal to its area.  Definition II views the particle positions as a point process, while definition I views the particles as defining a heterogeneous medium \cite{Zachary1}.

Defining the Fourier transform,
\begin{equation}
\phi_\mathbf{q}=\int_Vd^2r\,\mathrm{e}^{i\mathbf{q}\cdot\mathbf{r}}\phi(\mathbf{r}),
\end{equation}
fluctuations in the packing fraction at wavevector $\mathbf{q}$ are given by,
\begin{equation}
\chi(\mathbf{q})\equiv\frac{1}{V}\langle \phi_\mathbf{q}\phi_{-\mathbf{q}}\rangle.
\end{equation}
The signature of hyperuniformity is then
\begin{equation}
\chi(\mathbf{q})\sim |\mathbf{q}|\quad\mathrm{as}\quad|\mathbf{q}|\to 0,
\end{equation}
whereas $\chi(\mathbf{q}\to 0)\to$ constant if the system is not hyperuniform.

Note, using definition II of Eq.~(\ref{edefII})  we have,
\begin{equation}
\phi_\mathbf{q}=\sum_i v_i\mathrm{e}^{i\mathbf{q}\cdot\mathbf{r}_i},
\end{equation}
whereas using definition I of Eq.~(\ref{edefI}) we have,
\begin{equation}
\phi_\mathbf{q}=\sum_i \Delta_{i\mathbf{q}}\mathrm{e}^{i\mathbf{q}\cdot\mathbf{r}_i}
\label{ephiq}
\end{equation}
where $\Delta_{i\mathbf{q}}$ is the Fourier transform of $\Delta_i(\mathbf{r})$.  Since $\Delta_{i\mathbf{q}}\to v_i$ as $|\mathbf{q}|\to 0$, the two definitions of Eq.~(\ref{edefI}) and (\ref{edefII}) must give the same $\chi(\mathbf{q})$ in the limit $|\mathbf{q}|\to 0$, hence both are in principle good measures for hyperuniformity.

Note, for circular disks in two dimensions, $\Delta_{i\mathbf{q}}$ depends only on the magnitude $|\mathbf{q}|$ and is given by,
\begin{equation}
\Delta_{i\mathbf{q}}=v_if(|\mathbf{q}|d_i/2),\quad\mathrm\quad f(y)=\frac{2}{y^2}\int_0^ydx\,xJ_0(x).
\label{efy}
\end{equation}
Here $d_i$ is the diameter of the particle $i$ and $J_0(x)$ is the Bessel function of the first kind.

We will also consider another wavevector dependent measure of hyperuniformity, as introduced by Berthier et al.  \cite{Berthier}, the thermal compressibility $\chi_T(\mathbf{q})$ defined by,
\begin{equation}
\mathrm{definition\,III:}\qquad [n T\chi_T(\mathbf{q})]^{-1}=\sum_{s,s^\prime}x_sS^{-1}_{ss^\prime}(\mathbf{q})x_{s^\prime}.
\label{chiT}
\end{equation}
Here $n=N/V$ is the particle density, $s$ and $s^\prime$ label distinct species of particles of given diameter $d_s$, $x_s=N_s/N$ is the global concentration of species $s$, and $S^{-1}_{ss^\prime}(\mathbf{q})$ is the inverse of the matrix,
\begin{equation}
S_{ss^\prime}(\mathbf{q})\equiv\frac{1}{N}\langle n_{s\mathbf{q}}n_{s^\prime -\mathbf{q}}\rangle,
\end{equation}
where $n_{s\mathbf{q}}$ is the Fourier transform of the particle density of species $s$ alone.  
The quantity $\chi_T(\mathbf{q})$ in Eq.~(\ref{chiT}) is derived as the compressibility of a polydisperse liquid of particles in {\em thermal equilibrium} at temperature $T$.  For our {\em nonequilibrium} athermal system, in which fluctuations from configuration to configuration are induced by our rapid quench protocol rather than a finite temperature, the physical interpretation of $\chi_T(\mathbf{q})$ as a compressibility is unclear; nevertheless the right hand side of Eq.~(\ref{chiT}) is an interesting measure of density fluctuations, and so we will compute it for the sake of comparison.

\begin{figure}[h]
\begin{center}
\includegraphics[width=2.3in]{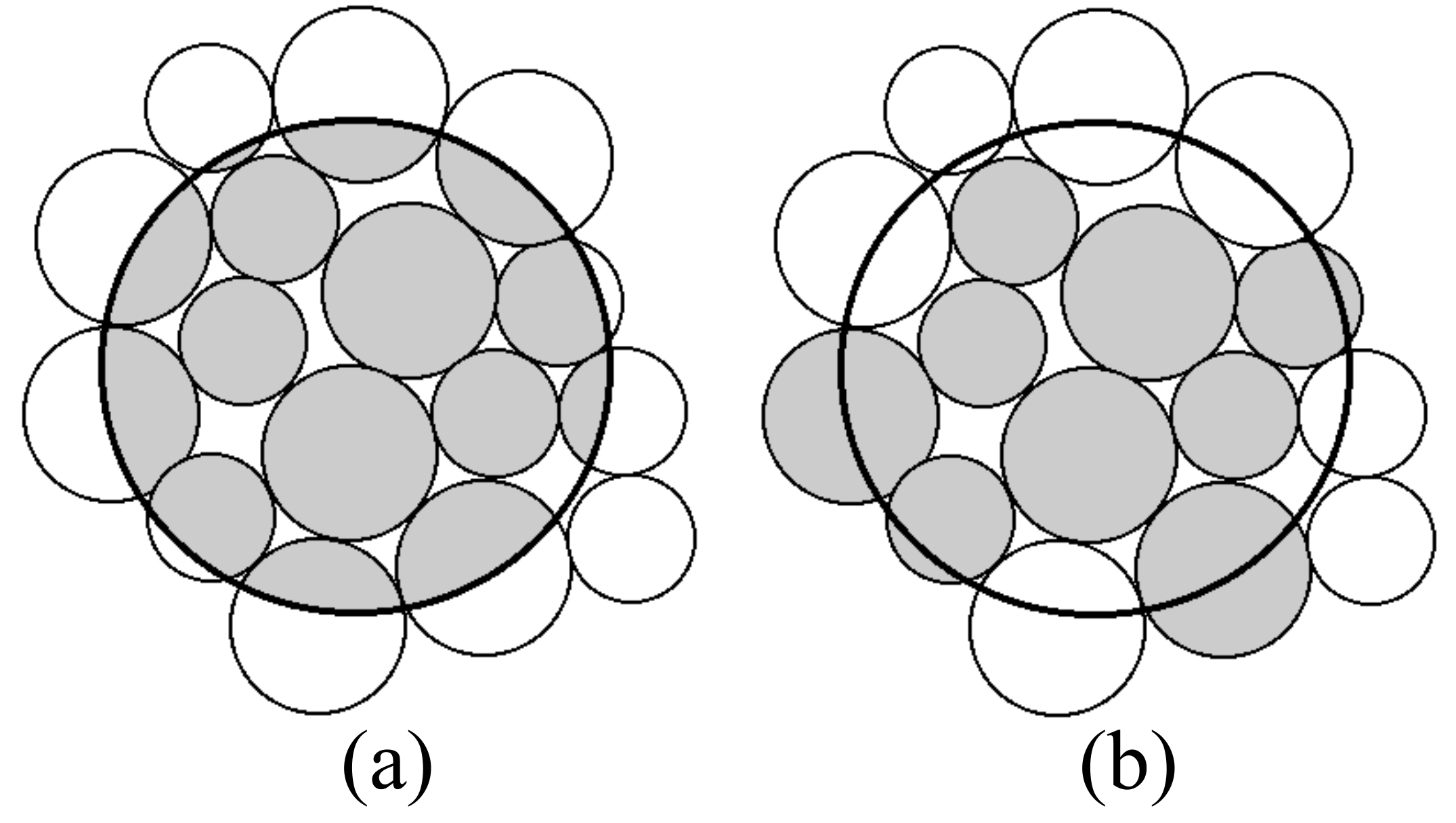}
\caption{Shaded volume represents the volume that contributes to $\phi_R$ according to (a) the definition I of $\phi(\mathbf{r})$ in Eq.~(\ref{edefI}), and (b) the definition II of $\phi(\mathbf{r})$ in Eq.~(\ref{edefII}).
}
\label{f1}
\end{center}
\end{figure}

We will also consider hyperuniformity as measured in real-space by computing the fluctuations of $\phi(\mathbf{r})$ over a circular window of radius $R$.  Place a circle of radius $R$ at a random position within the system and denote this region as the volume $V_R$.  We can then define the average packing fraction on this region as,
\begin{equation}
\phi_R\equiv\frac{1}{\pi R^2}\int_{V_R}d^2r\,\phi(\mathbf{r}).
\label{ephiR}
\end{equation} 
The difference in $\phi_R$ between using definition I of Eq.~(\ref{edefI}) and definition II of Eq.~(\ref{edefII}) for $\phi(\mathbf{r})$ is then as illustrated by the sketch in Fig.~\ref{f1}.  In definition I we count all overlapping volume between particles and the volume $V_R$; particles which are not entirely contained within $V_R$ contribute only the overlapping fraction of their volume, as illustrated.  In definition II we count the entire volume of particles whose centers lie within the volume $V_R$; particles whose centers lie outside $V_R$ contribute nothing, even if they overlap $V_R$.

We then compute the variance,
\begin{equation}
\mathrm{var}(\phi_R)\equiv\langle\phi_R^2\rangle-\langle\phi_R\rangle^2.
\end{equation}
One can show that $\mathrm{var}(\phi_R)$ is related to $\chi(\mathbf{q})$ by
\begin{equation}
\mathrm{var}(\phi_R)= \frac{1}{(\pi R^2)^2V}\sum_{\mathbf{q}\ne 0}\,\chi(\mathbf{q})\Delta_{R\mathbf{q}}^2,
\label{varphi-chi}
\end{equation}
where $\Delta_{R\mathbf{q}}$ is the Fourier transform
of the indicator function $\Delta_R(\mathbf{r})$ for a circular volume of radius $R$, 
and the sum is over all $\mathbf{q}$ consistent with Lees-Edwards boundary conditions excluding $\mathbf{q}=0$ \cite{note2}.

When $\chi(\mathbf{q})\to$ constant as $|\mathbf{q}|\to 0$, as in a liquid, the above gives \cite{Zachary2,Zachary3} for the limiting large $R$ behavior in two dimensions, 
\begin{equation}
\mathrm{for\,liquid:}\quad\mathrm{var}(\phi_R)\sim \frac{c}{R^2}.
\label{evarphiliq}
\end{equation}  
For a hyperuniform system, with $\chi(\mathbf{q})\sim |\mathbf{q}|$ as $|\mathbf{q}|\to 0$, the limiting large $R$ behavior in two dimensions is \cite{Zachary2,Zachary3},
\begin{equation}
\mathrm{for\,hyperuniform:}\quad\mathrm{var}(\phi_R)\sim\frac{a+b\ln R}{R^3}.
\label{evarphihyper}
\end{equation}

Since the $|\mathbf{q}|\to 0$ limiting behavior of $\chi(\mathbf{q})$ must be the same for definitions I and II, we expect that the large $R$ limiting behavior of $\mathrm{var}(\phi_R)$ must in principle also be the same.  However, unlike what we will find for $\chi(\mathbf{q})$, we will find that for the system sizes and length scales we can simulate, $\mathrm{var}(\phi_R)$ vs $R$ behaves very differently for the two definitions of $\phi(\mathbf{r})$.

The relative merits of the wavevector-dependent method $\chi(\mathbf{q})$ compared to the real-space method $\mathrm{var}(\phi_R)$, for detecting hyperuniformity as applied to particle images from physical experiments, has recently been discussed in Ref.~\cite{Dreyfus}.

\section{Model}
\label{SecModel}

Our two dimensional system of $N$ particles is a bidisperse mixture of equal numbers of big and small circular, frictionless, disks with diameters $d_b$ and $d_s$ in the ratio $d_b/d_s=1.4$ \cite{OHern}.  Disks $i$ and $j$ interact only when they overlap, in which case they repel with a soft-core interaction potential,
\begin{equation}
{\cal V}_{ij}(r_{ij})=\left\{
\begin{array}{ll}
\frac{1}{\alpha}k_e(1-r_{ij}/d_{ij})^\alpha, & r_{ij}<d_{ij}\\[10pt]
0, & r_{ij}\ge d_{ij}.
\end{array}
\right.
\label{einteraction}
\end{equation}
Here $r_{ij}$ is the center-to-center distance between the particles, and $d_{ij}=(d_i+d_j)/2$ is the sum of their radii.  We will measure energy in units such that $k_e=1$, and length in units so that the small disk diameter $d_s=1$.  Unless otherwise stated, our results are for the harmonic interaction with $\alpha=2$.

\begin{figure}[h]
\begin{center}
\includegraphics[width=1.8in]{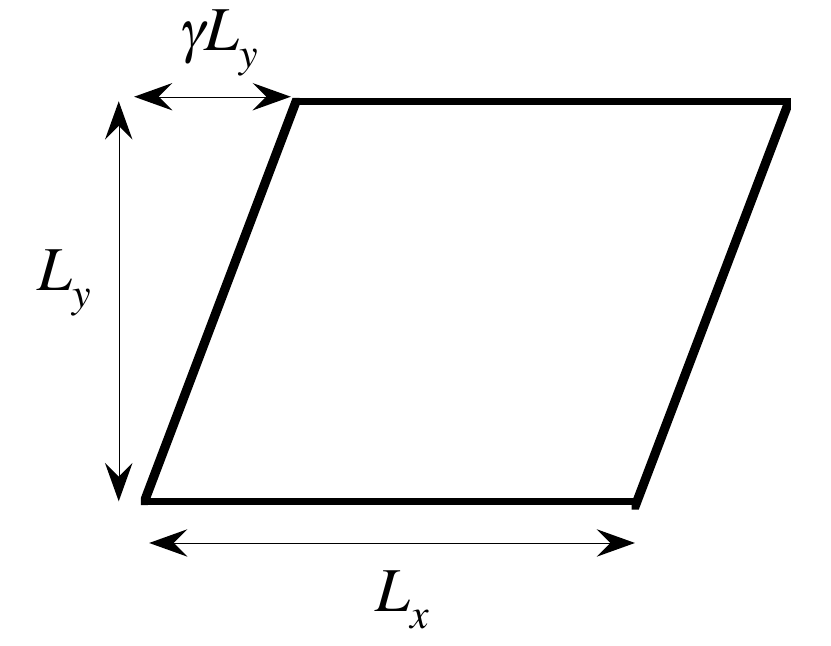}
\caption{Geometry of our system box.  $L_x$ and $L_y$ are the lengths in the $\mathbf{\hat x}$ and $\mathbf{\hat y}$ directions, and $\gamma$ is the skew ratio.  Lees-Edwards boundary conditions are used.
}
\label{f2}
\end{center}
\end{figure}

The geometry of our system box is characterized by three parameters, $L_x,L_y,\gamma$, as illustrated in Fig.~\ref{f2}. $L_x $ and $L_y$ are the lengths of the box in the $\mathbf{\hat x}$ and $\mathbf{\hat y}$ directions, while $\gamma$ is the skew ratio of the box.  We use Lees-Edwards boundary conditions \cite{LeesEdwards} to periodically repeat this box throughout all space.

In this work we consider only packings with an {\em isotropic} total stress tensor $\Sigma_{\alpha\beta}$,
\begin{equation}
\Sigma_{\alpha\beta}=\Gamma_N\delta_{\alpha\beta}, \quad \mathrm{where}\quad\Gamma_N=pV,
\end{equation}
$p$ is the system pressure, and $V=L_xL_y$ is the total system volume.   Here $\alpha,\beta$ denote the spatial coordinate directions $x,y$.  

To construct such isotropic packings, in which the shear stress vanishes, we use a scheme in which we vary the box parameters $L_x,L_y$ and $\gamma$ as we search for mechanically stable states \cite{Dagois}.
We introduce \cite{WuTeitel} a modified energy function $\tilde U$ that depends on the particle positions $\{\mathbf{r}_i\}$, as well as $L_x,L_y,\gamma$,
\begin{equation}
\tilde U\equiv U+\Gamma_N(\ln L_x +\ln L_y),\quad U\equiv\sum_{i<j}{\cal V}_{ij}(r_{ij}).
\end{equation}

Noting that the interaction energy $U$ depends implicitly on the box parameters $L_x,L_y,\gamma$ via the boundary conditions, we get the relations,
\begin{equation}
\begin{aligned}
L_x\frac{\partial U}{\partial L_x}=-\Sigma_{xx}+\gamma\Sigma_{xy},&\quad\frac{\partial U}{\partial \gamma}=-\Sigma_{xy},\\ 
L_y\frac{\partial U}{\partial L_y}=-\Sigma_{yy}-\gamma\Sigma_{xy}.&
\end{aligned}
\end{equation}

Starting from an initial configuration of randomly positioned particles in a square box ($L_x=L_y,\gamma=0$) at packing fraction $\phi_\mathrm{init}=0.84$, and fixing a target value of $\Gamma_N$, we then minimize $\tilde U$
with respect to both particle positions and box parameters.  
Our minimization can be considered as a rapid quench from infinite to zero temperature, keeping the final total system stress fixed.
The resulting local minimum of $\tilde U$
gives a mechanically stable configuration with force balance on each particle and a total stress tensor that satisfies 
\begin{equation}
\Sigma_{xx}=\Sigma_{yy}=\Gamma_N, \quad \Sigma_{xy}=0.  
\end{equation}
For minimization we use the Polak-Ribiere conjugate gradient algorithm \cite{NR}.  We consider the minimization converged when we satisfy the condition $(\tilde U_i - \tilde U_{i+50})/\tilde U_{i+50} < \varepsilon=10^{-10}$, where $\tilde U_i$ is the value at the $i$th step of the minimization.  Tests that our procedure gives well minimized configurations are discussed in the Appendix.
Our results at each value of $\Gamma_N$ are averaged over 1000--10000 (depending on the system size) independently generated isotropic configurations.  Configurations are generated independently at each value of $\Gamma_N$.

\section{Results}
\label{secResults}

We simulate our system for a range of total system stresses $\Gamma_N$ spanning just over two orders of magnitude.  It will be convenient to parameterize our configurations by the intensive quantity $\tilde p\equiv\Gamma_N/N$, the total stress per particle; $\tilde p$ is related to the ordinary pressure $p$ by $\tilde p = p(V/N)$.
We have considered four different system sizes, $N=8192$, $16384$, $32768$ and $65536$, each for equal values of $\tilde p =0.0001373$ to $\tilde p=0.0183105$.
We use large systems in two dimensions so as to be able to probe small wavevectors $q$, and so test for hyperuniformity on long length scales.

\subsection{Global quantities}
\label{sGlobal}

Before considering the behavior of local packing fraction fluctuations, we first consider several global properties of the system in order to establish where our systems lie with respect to the jamming transition.  For our model, a detailed finite-size-scaling analysis \cite{VagbergFSS} found that a rapid quench from random positions at {\em fixed packing fraction $\phi$}  gave a jamming fraction of $\phi_J=0.84177$.  However, since it is established \cite{Pinaki,VagbergGlass,Schreck,Krzakala,Parisi} that the jamming fraction $\phi_J$ of mechanically stable configurations can depend on the specific protocol used to produce those configurations, there is no guarantee that $\phi_J$ for rapid quenching to {\em constant stress $\tilde p$} necessarily results in the same exact value of $\phi_J$.

\begin{figure}[h]
\begin{center}
\includegraphics[width=3.4in]{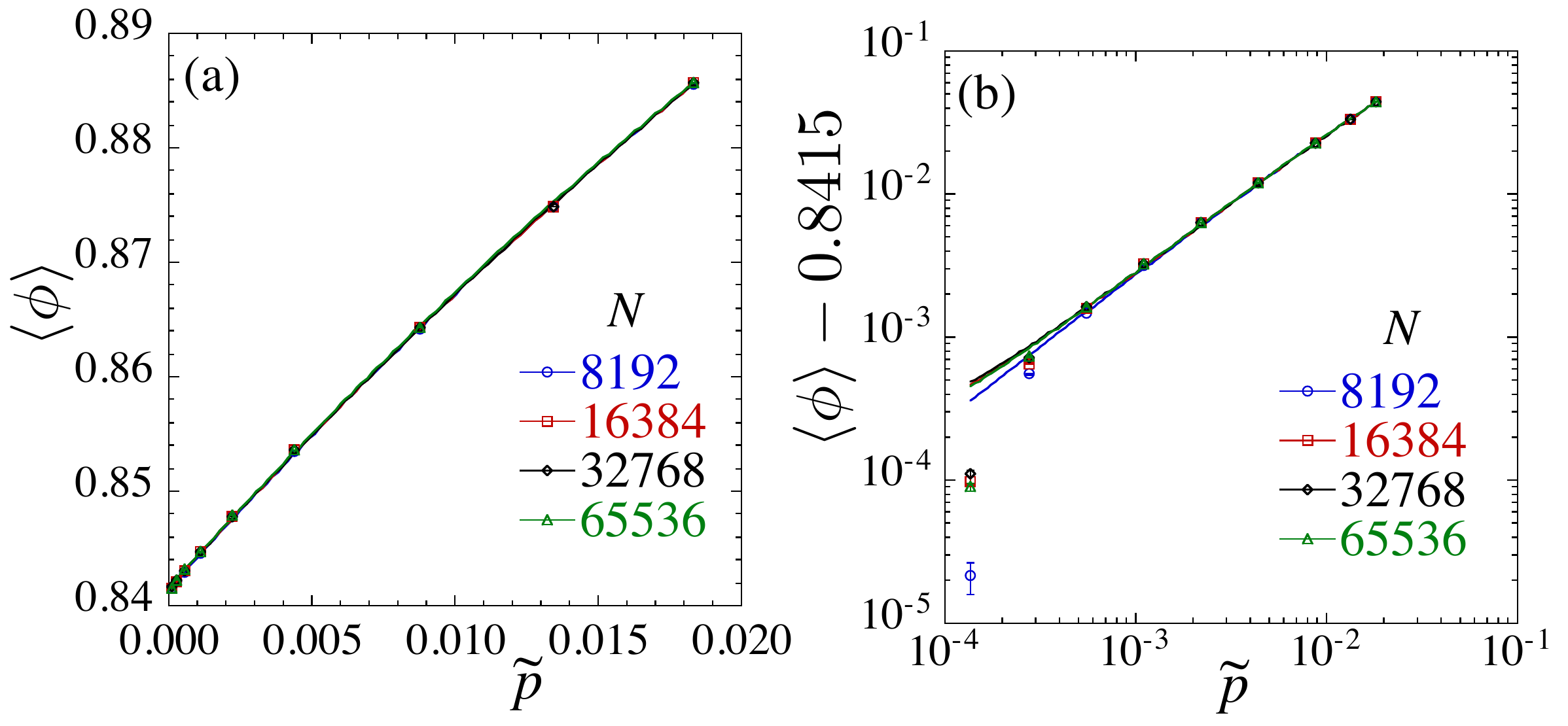}
\caption{(color online) Average global packing fraction $\langle\phi\rangle$ vs stress per particle $\tilde p$, for systems with $N=8192$ to $65536$ particles on (a) linear-linear, and (b) log-log scales.  Solid lines are a fit to a quadratic function $\langle\phi\rangle=\phi_J+a_1\tilde p+a_2\tilde p^2$.  
}
\label{phi-vs-p}
\end{center}
\end{figure}

Since our minimization procedure varies the box lengths $L_x$ and $L_y$ to achieve the desired global stress $\Gamma_N$, different configurations at a common fixed $\Gamma_N$ may have slightly different volumes (see Appendix), and hence different global packing fractions.
In Fig.~\ref{phi-vs-p} we plot the average global packing fraction $\langle\phi\rangle$ vs the stress per particle $\tilde p$, for systems with $N=8192$ to $65536$ particles.  Panel (a) shows the results on a linear-linear scale, where it appears that finite size effects are negligible.  It has been predicted \cite{OHern} that, for our harmonic interaction of Eq.~(\ref{einteraction}), pressure scales linearly with packing fraction. In our data, however, we see a small but clear curvature at larger $\tilde p$.  We thus fit (solid lines in Fig.~\ref{phi-vs-p}) our results to $\langle\phi\rangle = \phi_J + a_1\tilde p+a_2\tilde p^2$, regarding the quadratic term as a correction to scaling.  This fit gives $\phi_J=0.84159\pm 0.00002$, with the error representing the variation in values obtained for the different system sizes.

However, to examine more closely the points at the smallest $\tilde p$, in Fig.~\ref{phi-vs-p}b we plot $\langle\phi\rangle-0.8415$ vs $\tilde p$ on a log-log scale \cite{note3}.  We now see a definite finite size effect in the results for the two smallest $\tilde p$ values, and that these lie noticeably below the fitted quadratic curve (solid line).  Our above estimate of $\phi_J$ should therefore be taken with some caution.  A more accurate determination of $\phi_J$, as well as the power-law dependence between $\tilde p$ and $(\phi-\phi_J)$, should take into account these finite size effects.  Such an analysis is outside the scope of the present work.

\begin{figure}[h]
\begin{center}
\includegraphics[width=3.4in]{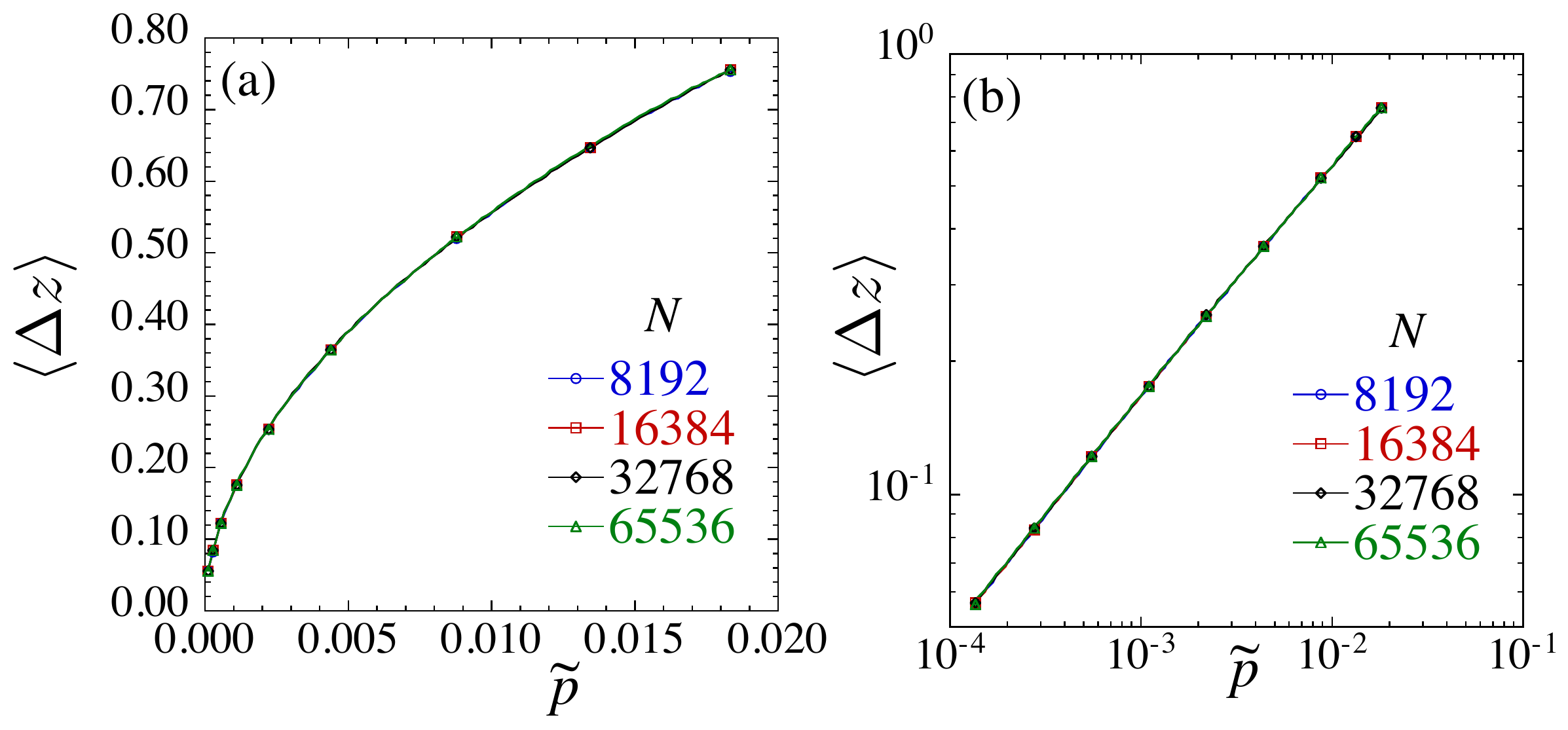}
\caption{(color online) Average excess contact number $\langle\Delta z\rangle=\langle z\rangle-z_\mathrm{iso}$ vs stress per particle $\tilde p$, for systems with $N=8192$ to $65536$ particles on (a) linear-linear, and (b) log-log scales.  Solid lines are a fit to $\langle\Delta z\rangle=\tilde p^{x}(a_0+a_1\tilde p+a_2\tilde p^2+a_2\tilde p^3)$, where we find $x\approx 0.546$.
}
\label{Dz-vs-p}
\end{center}
\end{figure}

The jamming transition of frictionless particles is well determined by the {\em isostatic} condition \cite{Liu+Nagel,OHern, vanHecke}, where the number of constraints on the particles exactly equals the number of degrees of freedom.  For frictionless spherical particles this condition requires that
the average number of contacts $\langle z\rangle$ for a given particle is equal to twice the dimensionality of the system; for two dimensions, $z_\mathrm{iso}=4$.  Numerically, this condition is found to hold quite precisely provided one first excludes from the system ``rattler" particles \cite{OHern}.  A rattler is any particle which is not at a strict local energy minimum, but may move without cost in energy in one or more directions.  
To locate the rattlers in our two dimensional system we loop recursively through all our particles removing any particle with less than three contacts;  i.e., after an initial pass in which such particles are removed, we loop again through the remaining particles and remove any that now have less than three contacts, repeating this procedure until no additional particles are removed.  The total number of removed particles is then the number of rattlers.
Removing such rattlers and computing the resulting average $\langle z\rangle$ of the remaining particles,
in Fig.~\ref{Dz-vs-p} we plot $\langle\Delta z\rangle\equiv \langle z\rangle - z_\mathrm{iso}$, vs $\tilde p$, for system sizes $N=8192$ to $65536$.  Panel (a) shows our results on a linear-linear scale, while panel (b) shows a log-log scale.  In this case, as has been noted previously \cite{Goodrich}, finite size effects are truly negligible for the range of $\tilde p$ and $N$ considered here. For the harmonic interaction used here, theoretical arguments \cite{Wyart2} predict $\langle \Delta z\rangle\sim p^{1/2}$ close to the jamming transition.  Since our values of $\tilde p$ extend moderately above jamming, $(\langle\phi\rangle_\mathrm{max}-\phi_J)/\phi_J\approx 0.05$, we fit our data to the form $\langle\Delta z\rangle=\tilde p^x(a_0+a_1\tilde p+a_2\tilde p^2+a_2\tilde p^3)$, where the polynomial factor is an empirical form to account for corrections to scaling when not sufficiently close to jamming.  We find the value $x\approx 0.546\pm 0.001$, with the error representing the variation in values obtained for the different system sizes.  A more careful scaling analysis, going to lower stresses $\tilde p$ closer to jamming, is desirable before concluding the exponent is truly $x>1/2$.  However we may note that a recent reanalysis \cite{Andrea} of the data of Ref.~\cite{Goodrich} has similarly found values of $x>1/2$ in both two and three dimensions.

\begin{figure}[h]
\begin{center}
\includegraphics[width=3.4in]{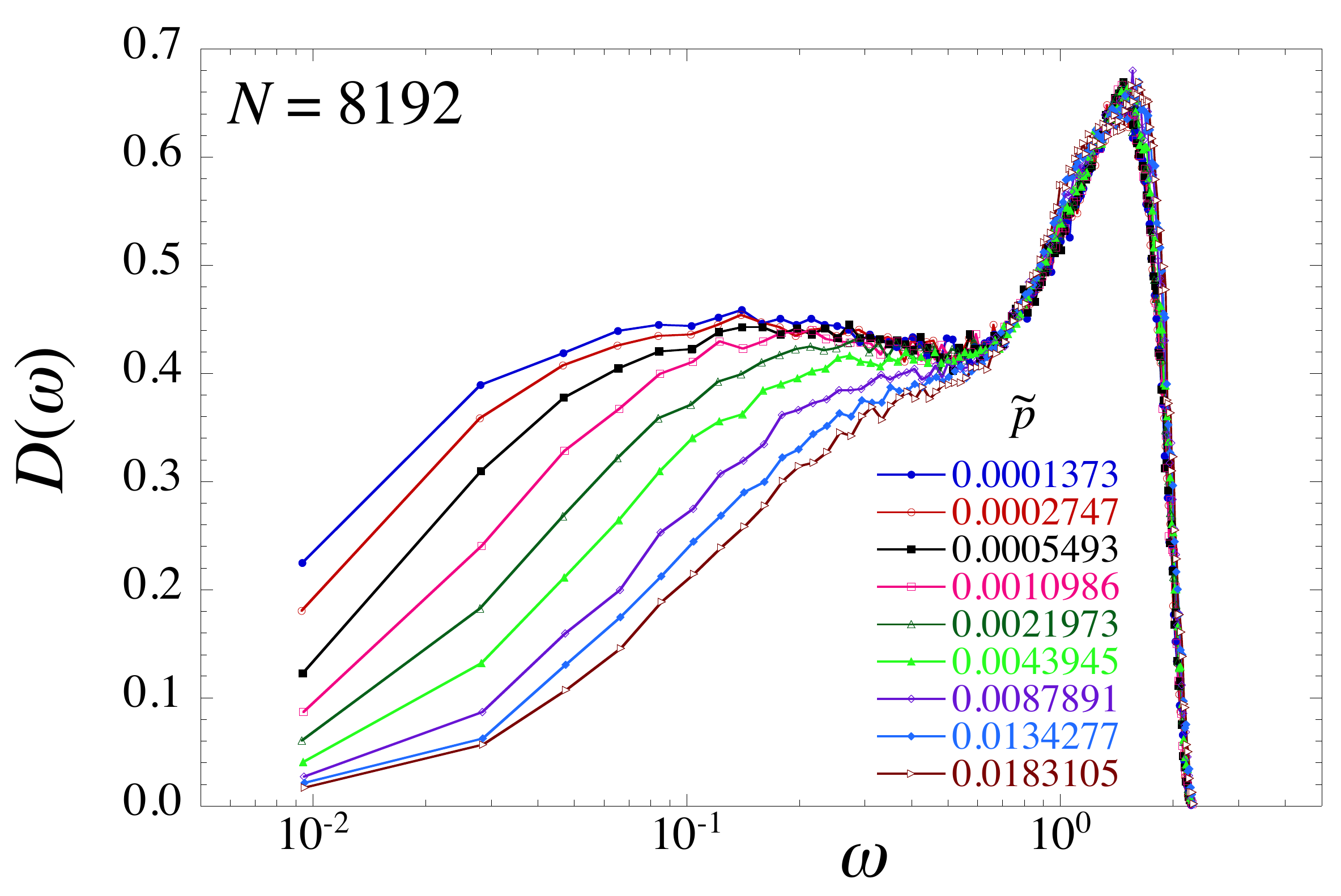}
\caption{(color online) Density of frequencies of small elastic vibrations $D(\omega)$ vs $\omega$.  Results are shown for a system of $N=8192$ particles for a range of stress per particle $\tilde p=0.0001373$ to $0.0183105$.  Each curve is an average over 3 independent energy minimized configurations.
}
\label{Dw-log}
\end{center}
\end{figure}

As a final measure of the global properties of our systems we consider the density of states $D(\omega)$ of the dynamical matrix of our minimized configurations \cite{OHern,Wyart2,Wyart}. Expanding the interaction energy $U(\{\mathbf{r}_i\})$ to second order in small particle displacements about the energy minimized configuration defines the dynamical matrix.  The eigenvalues $\lambda$ of that matrix, and corresponding eigenvectors, determine the response of the system to vanishingly small elastic perturbations.  Following convention and assuming Newtonian equations of motion for the response to such perturbations, the eigenvalues $\lambda$ are related to the frequencies of the normal modes of vibration $\omega$  by $\lambda=\omega^2$.  In Fig.~\ref{Dw-log} we plot the density of such frequencies $D(\omega)$ vs $\omega$ on a linear-log scale, for a stress per particle ranging from $\tilde p=0.0001373$ to $0.0183105$.  Because of the numerical difficulty of computing the eigenvalue spectrum for large matrices, we show results only for our smallest system size with $N=8192$ particles; curves at each $\tilde p$ are averaged over 3 independent configurations. We see clearly the plateau at small $\omega$, often referred to as the ``boson peak" \cite{OHern}, that shows the excess of low frequency modes characteristic of a marginally stable solid.  As $\tilde p$ decreases, the low frequency edge of the plateau, $\omega^*$, moves steadily to lower values and presumably vanishes as $\tilde p\to 0$ \cite{OHern}.  Our range of stress $\tilde p$ is thus clearly in the region where marginal stability is characterizing the structure of the packing out to ever increasing length scales as $\tilde p$ decreases.

\subsection{Wavevector-dependent fluctuations}

We now consider the fluctuations of the system at finite wavevectors $\mathbf{q}$.  
For the system geometry of Fig.~\ref{f2}, the wavevectors allowed by the Lees-Edwards boundary conditions have the form, $\mathbf{q}=m_1\mathbf{b}_1+m_2\mathbf{b}_2$, where $m_1$ and $m_2$ are integers, and the basis vectors are $\mathbf{b}_1=(2\pi/L_x)(\mathbf{\hat x}-\gamma\mathbf{\hat y})$ and $\mathbf{b}_2=(2\pi/L_y)\mathbf{\hat y}$.
For simplicity we will look at wavevectors oriented in the $\mathbf{\hat y}$ direction, i.e. $\mathbf{q}=m\mathbf{b}_2=q\mathbf{\hat y}$, with $q=2\pi m/L_y$ for integer $m$ \cite{qdirection}.  Because each different configuration may have a slightly different value of $L_y$, since $L_y$ is a free variable determined by the targeted value of $\Gamma_N$, we average data points at common values of $m$; however the variation in $L_y$ over different configurations, while finite, is in practice negligible for the large system sizes we consider here (see Appendix).

In Fig.~\ref{S-vs-q} we plot the structure function $S(q\mathbf{\hat y})$, that measures fluctuations of particle density, vs $q$ for a system of $N=32768$ particles for a range of stresses $\tilde p$.  As expected, we see that $S(q\mathbf{\hat y})$ saturates to a finite constant as $q$ decreases, for all $\tilde p$.  

\begin{figure}[h]
\begin{center}
\includegraphics[width=3.4in]{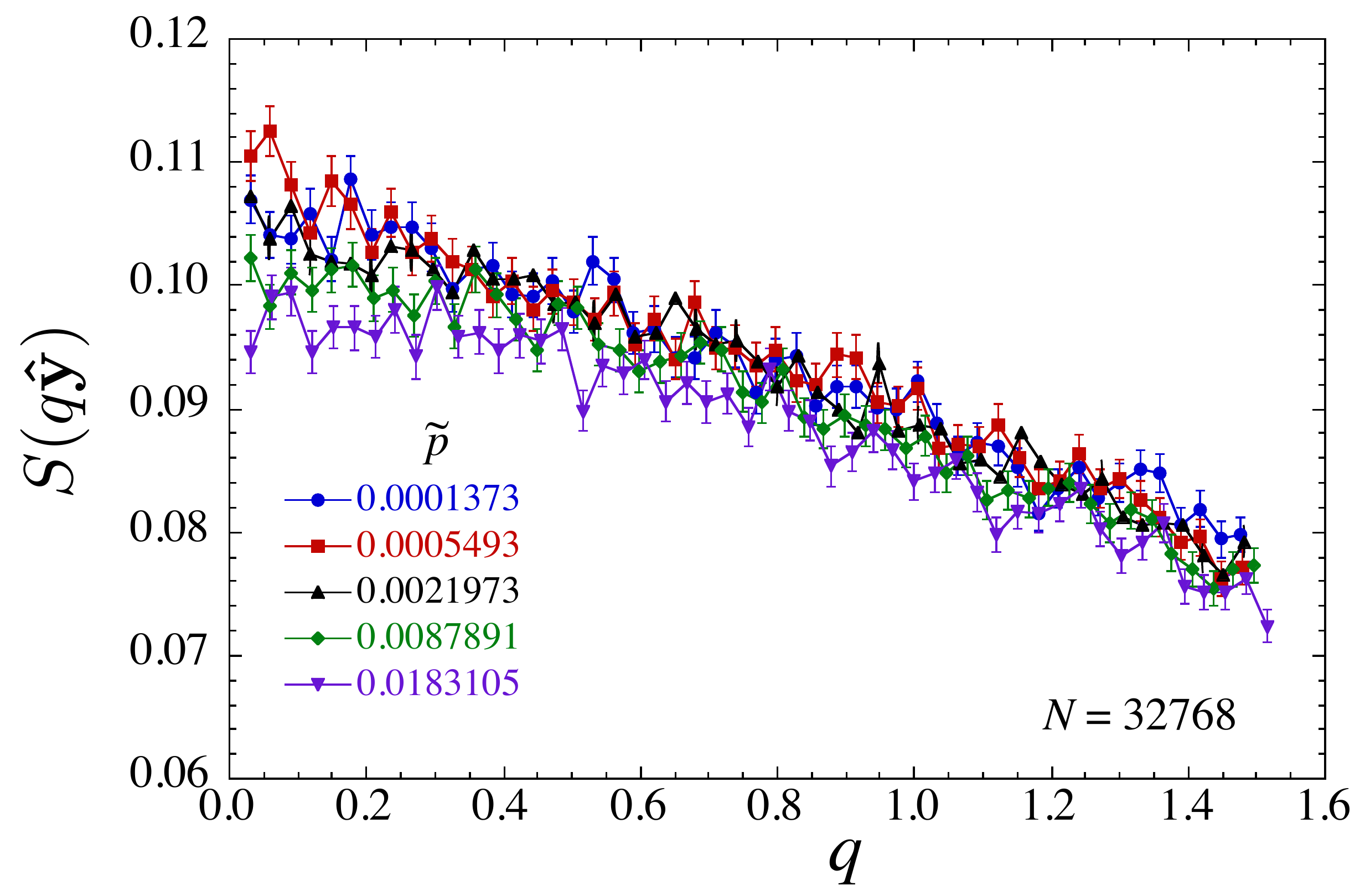}
\caption{(color online) Structure function $S(q\mathbf{\hat y})$ vs $q$, giving the fluctuation in the density of particles at wavevector $q\mathbf{\hat y}$.  Results are shown for a system with $N=32768$ particles, for several different values of the stress per particle, $\tilde p=0.0001373$ to $0.0183105$.  $S(q\mathbf{\hat y})$ approaches a constant as $q\to 0$.  The error bars shown in the figure represent one standard deviation of estimated statistical error.
}
\label{S-vs-q}
\end{center}
\end{figure} 

\begin{figure}[h]
\begin{center}
\includegraphics[width=3.4in]{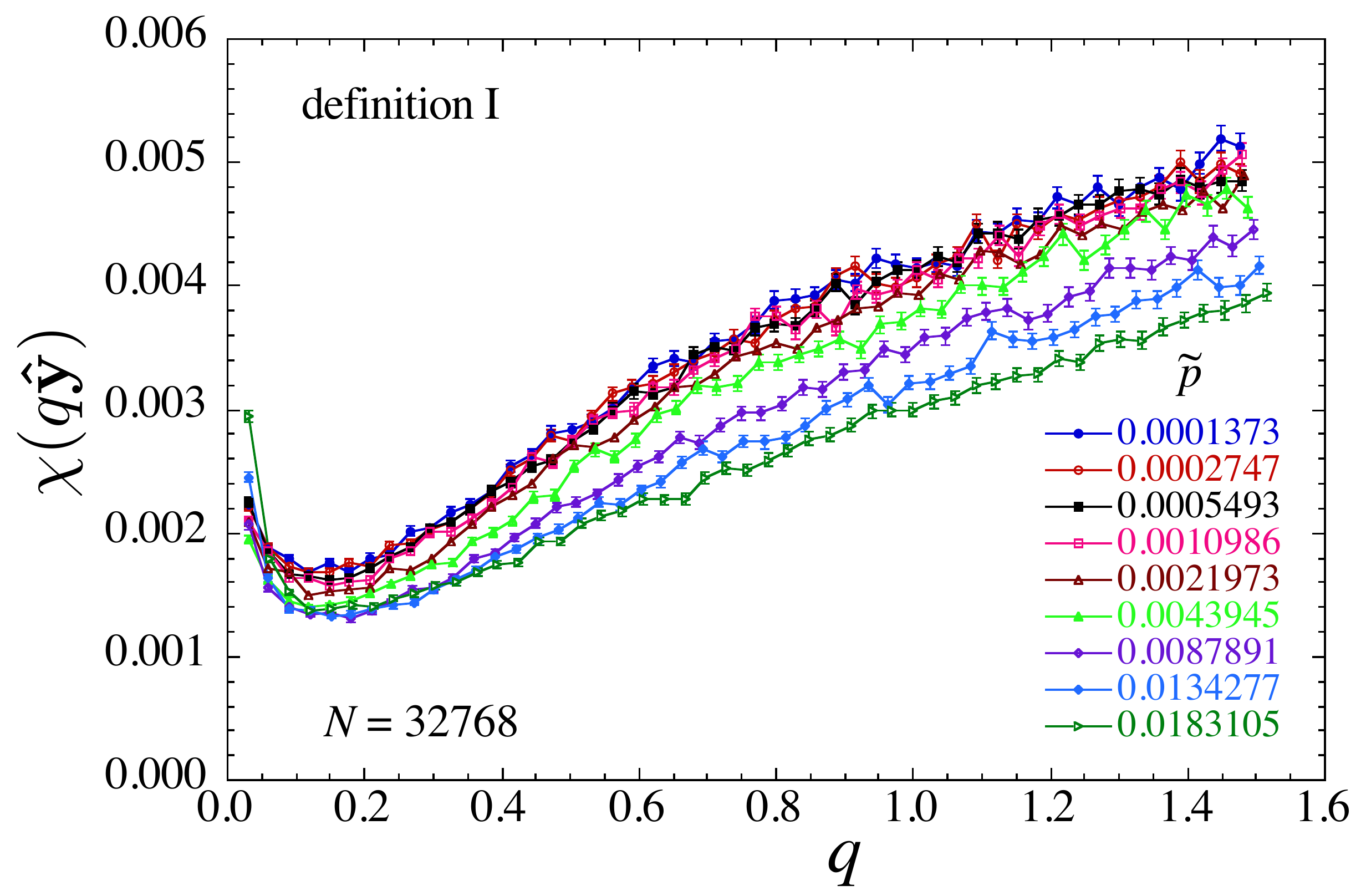}
\caption{ (color online) Fluctuation in packing fraction $\chi(q\mathbf{\hat y})$ vs $q$, using the definition I of Eq.~(\ref{edefI}) for the local packing fraction $\phi(\mathbf{r})$.  Results are shown for a system with $N=32768$ particles, for several different values of the stress per particle, $\tilde p=0.0001373$ to $0.0183105$. As $q$ decreases, we find that $\chi(q\mathbf{\hat y})$ decreases roughly linearly, but then reaches a minimum below which it increases, thus implying the absence of hyperuniformity on large length scales.  As $\tilde p$ decreases, the curves of $\chi(q\mathbf{\hat y})$ approach a common limiting curve.  The error bars shown in the figure represent one standard deviation of estimated statistical error.
}
\label{chiI-vs-q}
\end{center}
\end{figure}

In Fig.~\ref{chiI-vs-q} we show our results for the fluctuations of the local packing fraction, plotting $\chi(q\mathbf{\hat y})$ vs $q$, where we have used definition I of Eq.~(\ref{edefI}) for the local packing fraction $\phi(\mathbf{r})$.  We show results for a system of $N=32768$ particles for a range of different stresses $\tilde p$.  We see that as $q$ decreases, $\chi(q\mathbf{\hat y})$ decreases roughly linearly in $q$ as was observed previously.  However when $q$ gets sufficiently small, $\chi$ reaches a finite minimum at a $q^*$, and then {\em increases} as $q\to 0$, rather than vanishing as expected for a hyperuniform system.  Note that the limiting value $\chi(q\mathbf{\hat y}\to 0)$ is increasing as $\tilde p$ increases.

The variation of $q^*$ with $\tilde p$ is quite small.  As $\tilde p$ decreases, $q^*$ decreases slightly, but at sufficiently small $\tilde p$, the data at different $\tilde p$ appear to be approaching a common curve, with a common limiting value of $q^*\approx 0.15$.  We thus conclude that as $\tilde p$ decreases, and one approaches the jamming transition, our configurations display hyperuniformity only out to a {\em finite} length scale $\ell^*\approx 2\pi/q^*\approx 42$.  This is the main result of this work.  

\begin{figure}[h]
\begin{center}
\includegraphics[width=3.4in]{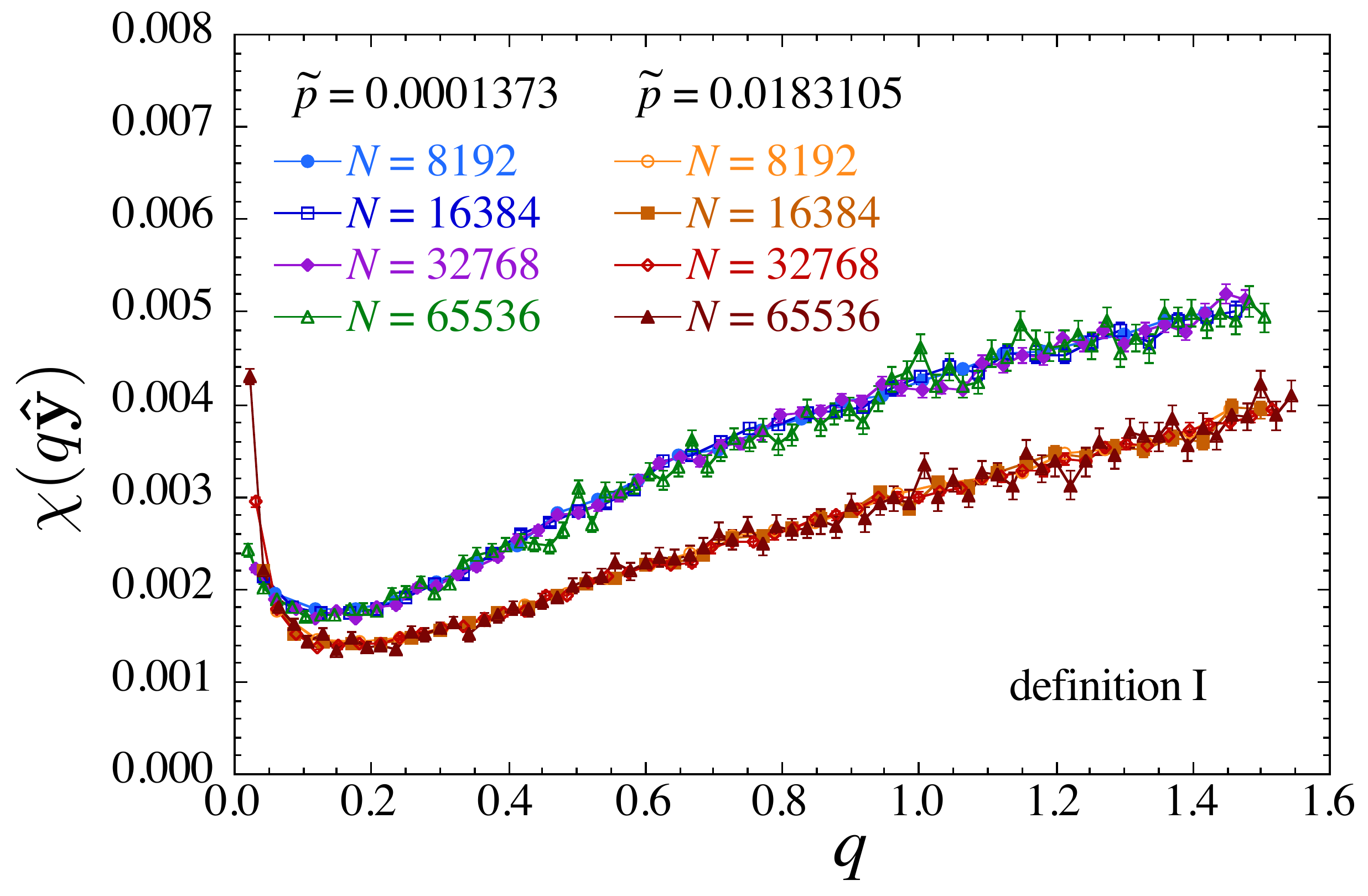}
\caption{(color online) Fluctuation in packing fraction $\chi(q\mathbf{\hat y})$ vs $q$, using the definition I of Eq.~(\ref{edefI}) for the local packing fraction $\phi(\mathbf{r})$.  Results are shown for several different system sizes, $N=8192$ to $65536$, for our smallest and largest values of the stress per particle, $\tilde p=0.0001373$ and $0.0183105$.  The error bars shown in the figure represent one standard deviation of estimated statistical error.
}
\label{chiI-vs-q-FS}
\end{center}
\end{figure}

One may question whether our observed behavior of $\chi(q\mathbf{\hat y})$ at small $q$ is not some artifact of our numerical procedure.  In the Appendix we show a careful analysis that this behavior is not an artifact of an insufficiently converged minimization procedure. Another possibility might be that it is a finite size effect.  In Fig.~\ref{chiI-vs-q-FS} we therefore plot $\chi(q\mathbf{\hat y})$ vs $q$ (using definition I for $\phi(\mathbf{r})$) for several different system sizes, $N=8192$ to $65536$, for our smallest stress $\tilde p=0.0001373$ and for our largest stress $\tilde p=0.0183105$.  Apart from the fact that  in systems with larger $N$ we can measure down to smaller $q$ (since $q=2\pi m/L_y$), the  measured $\chi(q\mathbf{\hat y})$ is found to be completely independent of the system size.

\begin{figure}[h]
\begin{center}
\includegraphics[width=3.4in]{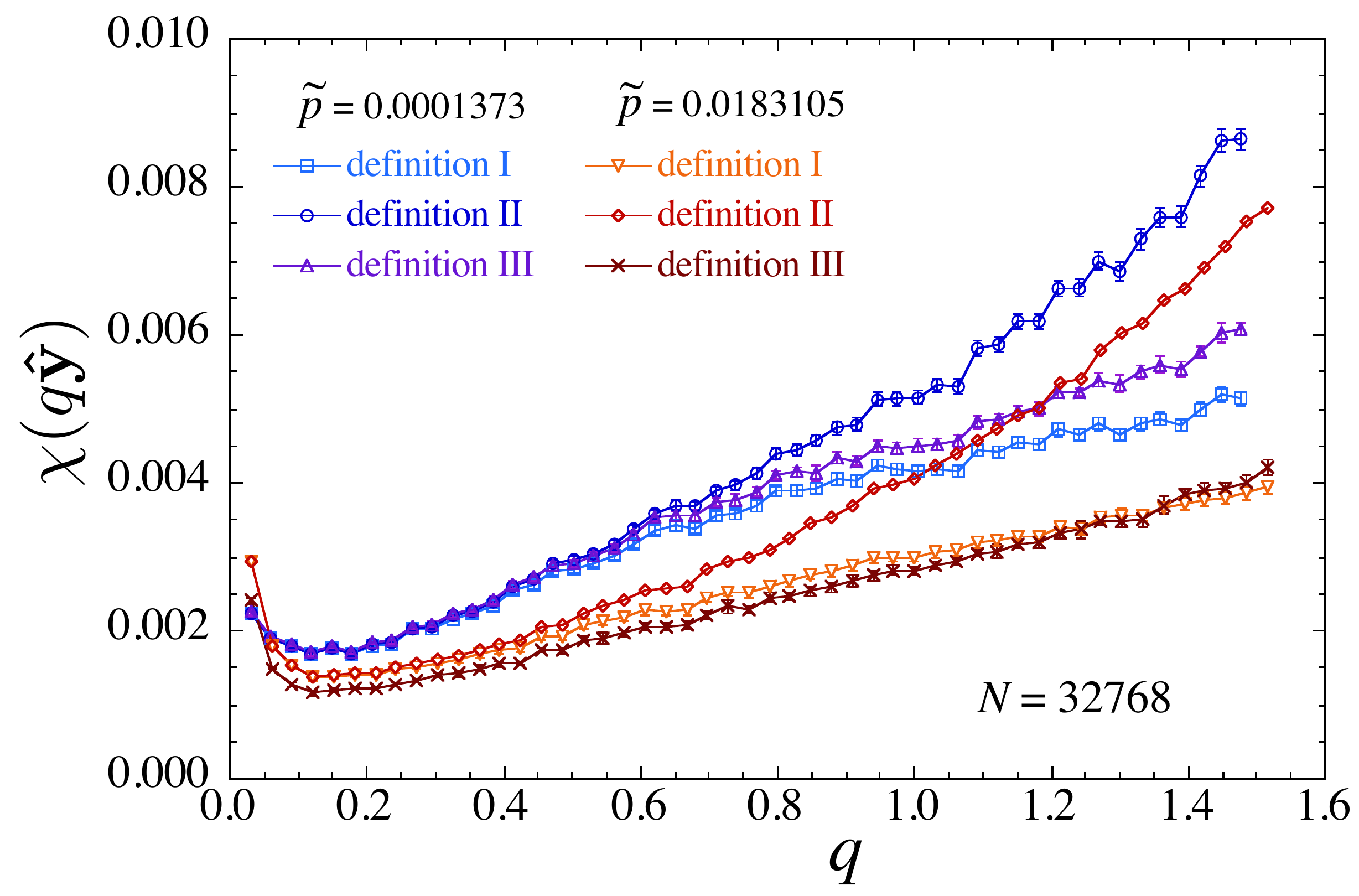}
\caption{(color online) Comparison of the different wavevector-dependent measures of hyperuniformity: ``definition I" and ``definition II" refer to the fluctuation of local packing fraction $\chi(\mathbf{q})$ with $\phi(\mathbf{r})$ determined from the definitions of Eq.~(\ref{edefI}) and Eq.~(\ref{edefII}) respectively, while ``definition III" refers to the quantity $nT\chi_T(\mathbf{q})$ defined in Eq.~(\ref{chiT}). Results are shown for a system with $N=32768$ particles, for our smallest and largest values of the stress per particle, $\tilde p=0.0001373$ and $0.0183105$.  The error bars shown in the figure represent one standard deviation of estimated statistical error.
}
\label{chi-1-2-B-vs-q}
\end{center}
\end{figure}

Finally we consider our two other wavevector-dependent measures of hyperuniformity, the fluctuation $\chi(\mathbf{q})$ using definition II of Eq.~(\ref{edefII}) for the local packing fraction $\phi(\mathbf{r})$, and the thermal compressibility $nT\chi_T(\mathbf{q})$ of Eq.~(\ref{chiT}), used by Berthier et al. \cite{Berthier}, which we denote as ``definition III."  In Fig.~\ref{chi-1-2-B-vs-q} we plot $\chi(q\mathbf{\hat y})$ vs $q$ for definitions I, II, and III, for a system with $N=32768$ particles at our smallest and largest values of $\tilde p$.  While these quantities all differ somewhat at the larger values of $q$, we see that definitions I and II become completely equal at smaller $q$, in particular about the minimum $q^*$, as should be expected from the discussion following Eq.~(\ref{ephiq}). Definition III for $nT\chi_T$ is completely equal to definitions I and II at small $q$ about the minimum $q^*$ at our lowest $\tilde p = 0.0001373$.  For the largest $\tilde p= 0.0183105$ we find a small deviation between $nT\chi_T$ and $\chi$ that persists to low $q$ at and below $q^*$; however the qualitative behavior remains the same.  We thus conclude that all three approaches lead to the same conclusion: that hyperuniformity extends only out to a finite length scale for our mechanically stable packings above the jamming transition, and that this length remains finite as the jamming transition is approached.

\subsection{Real space fluctuations}

In this section we consider the real space fluctuations of the local packing fraction, defined over a circular window of radius $R$, by computing the variance of $\phi_R$ as defined in Eq.~(\ref{ephiR}).  For each configuration we use several different, non-overlapping, circular windows at each given $R$.  When the diameter $2R$ is roughly equal to half the length of the system $L/2$, we take only a single window per configuration.

\begin{figure}[h]
\begin{center}
\includegraphics[width=3.4in]{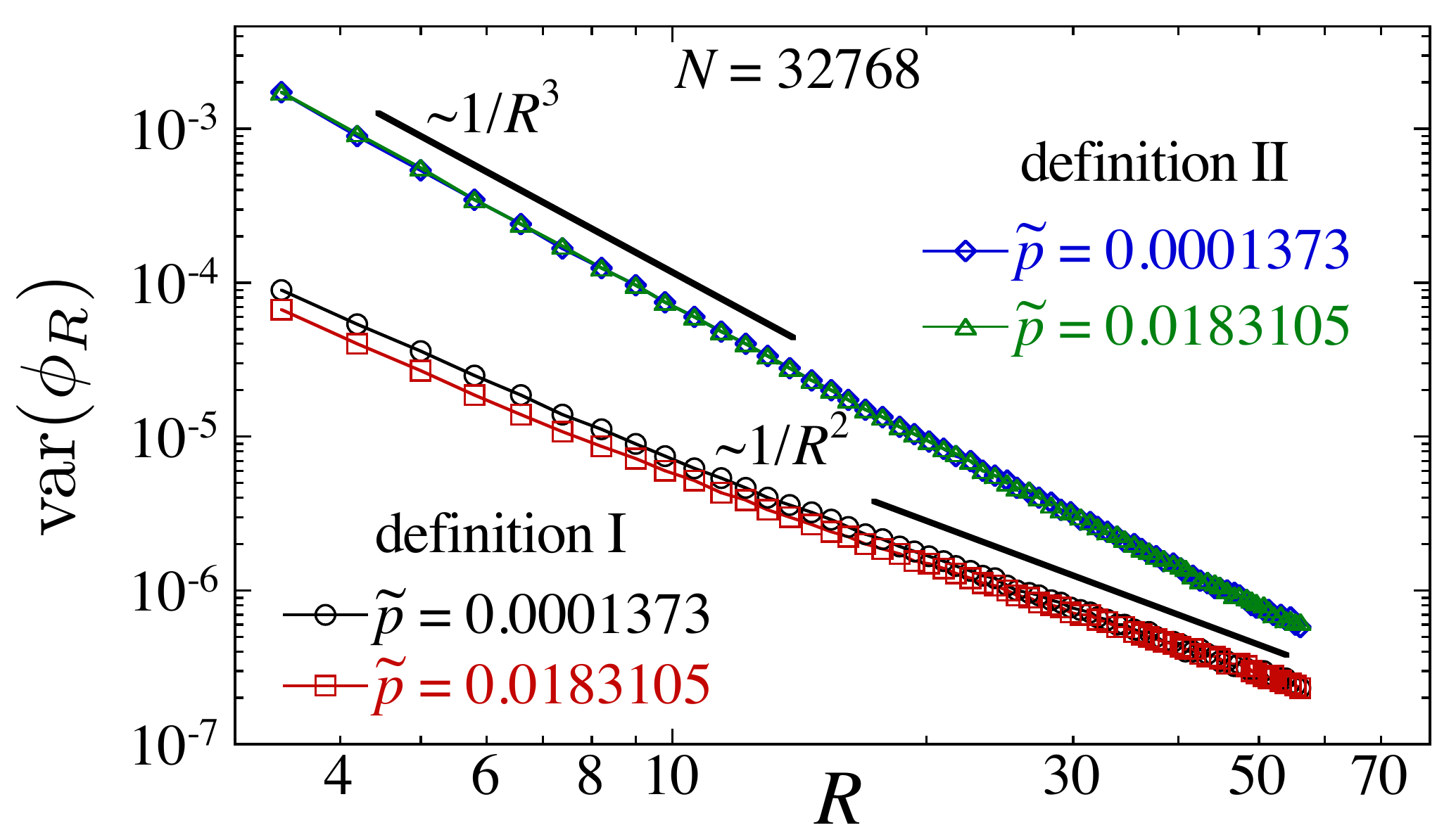}
\caption{(color online) Comparison of the fluctuation in local packing fraction, $\mathrm{var}(\phi_R)$ vs $R$, using definition I of Eq.~(\ref{edefI}) and definition II of Eq.~(\ref{edefII}) for the local packing fraction $\phi(\mathbf{r})$.  Results are shown for a system with $N=32768$ particles, for our smallest and largest values of the stress per particle, $\tilde p=0.0001373$ and $0.0183105$.  Thick solid lines represent the dependancies $1/R^2$ and $1/R^3$, as indicated.
}
\label{varphi-def-I-def-II}
\end{center}
\end{figure}

In Fig.~\ref{varphi-def-I-def-II} we plot $\mathrm{var}(\phi_R)$ vs $R$, comparing results from using definition I of Eq.~(\ref{edefI}) for the local packing fraction $\phi(\mathbf{r})$ with that of definition II of Eq.~(\ref{edefII}).  We show results for our smallest stress $\tilde p=0.00011373$ and our largest $\tilde p=0.0183105$, for a system with $N=32768$ particles.  Although the corresponding $\chi(q\mathbf{\hat y})$ for these two definitions were shown in Fig.~\ref{chi-1-2-B-vs-q} to be essentially identical at small $q$, we see a rather dramatic difference in the behaviors of the corresponding $\mathrm{var}(\phi_R)$ for the entire range of $R$ we study.  
As expected from Fig.~\ref{chi-1-2-B-vs-q}, the fluctuations for definition I are smaller than for definition II.  However the two definitions also appear to give different power-law dependencies for the decay of $\mathrm{var}(\phi_R)$ with $R$.  Definition II gives roughly a $1/R^3$ decay, while definition I seems to be closer to a $1/R^2$ decay at large length scales.  We will see below that the big difference in magnitude of $\mathrm{var}(\phi_R)$ comparing definition I and definition II, as well as the apparent difference in power-law decay, can be attributed to the contributions to $\mathrm{var}(\phi_R)$ from moderate to large $|\mathbf{q}|$ fluctuations (i.e. small length scale fluctuations), and that these higher $|\mathbf{q}|$ fluctuations are much larger for definition II.

\begin{figure}[h]
\begin{center}
\includegraphics[width=3.4in]{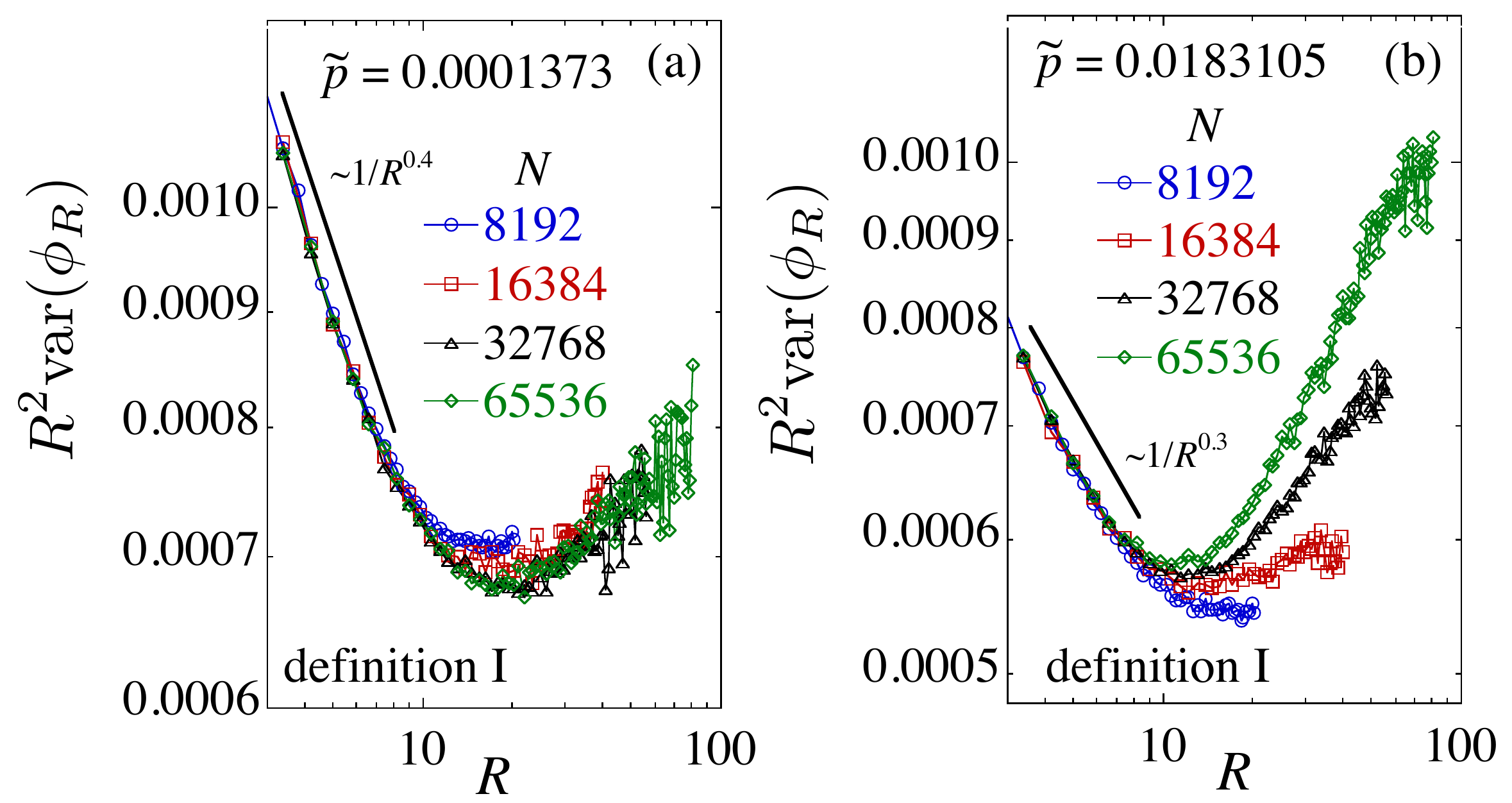}
\caption{(color online) $R^2\mathrm{var}(\phi_R)$ vs window radius $R$ for systems with $N=8192$ to $65536$ particles, at stress (a) $\tilde p=0.0001373$ and (b) $\tilde p=0.0183105$.  The power law decay at small $R$ is indicated.  Definition I of Eq.~(\ref{edefI}) for the local packing fraction $\phi(\mathbf{r})$ is used. 
}
\label{R2varphi-def-I}
\end{center}
\end{figure}

To examine this decay more closely, we consider $R^2\mathrm{var}(\phi_R)$, which according to Eqs.~(\ref{evarphiliq}) and (\ref{evarphihyper}) should approach a constant for a liquid-like system, and $(a+b\ln R)/R$ for a hyperuniform system.  In Fig.~\ref{R2varphi-def-I} we show $R^2\mathrm{var}(\phi_R)$ vs $R$ using definition I, for our smallest and largest stresses, $\tilde p=0.0001373$ and $\tilde p=0.0183105$, for several different system sizes from $N=8192$ to $65536$.  
At small $R$ we see that $R^2\mathrm{var}(\phi_R)$ decays as $R$ increases. A power-law fit to the small $R$ data in panel (a) gives a decay $\sim R^{-0.4}$, while in panel (b) we find $\sim R^{-0.3}$; it is not clear that these exponent values have any fundamental significance.
However as $R$ increases, this decay is cutoff at a length $R^*$ where $R^2\mathrm{var}(\phi_R)$ reaches a minimum.  Comparing panels (a) and (b) we see that $R^*$ decreases only slightly as $\tilde p$ increases over the two orders  of magnitude.  At the lowest stress, $R^*\approx 18$, corresponding to a window of diameter $2R^*=36$; this is
roughly consistent with the value of $\ell^*=2\pi/q^*\approx 42$ obtained from the minimum of $\chi(q\mathbf{\hat y})$ in Fig.~\ref{chiI-vs-q}.  

For $R>R^*$ we see that $R^2\mathrm{var}(\phi_R)$ increases, rather than saturating to a constant as might be expected.  This is the real space manifestation of the increase in $\chi(q\mathbf{\hat y})$ as $q$ decreases below $q^*$.  Whether $R^2\mathrm{var}(\phi_R)$ will continue to increase, or saturate to a constant, as $R$ increases further (i.e. whether $\chi(q\mathbf{\hat y})$ continues to increase or saturates to a constant as $q\to 0$) remains unclear.  The finite size dependence seen at large $R$ is another reflection of the increase in $\chi(q\mathbf{\hat y})$ as $q$ decreases below $q^*$.  From Eq.~(\ref{varphi-chi}) we have that $R^2\mathrm{var}(\phi_R)$ is related to the sum of $\chi(\mathbf{q})$ over all allowed wavevectors.  As $N$ increases, the smallest allowed $\mathbf{q}$ decreases ($q_\mathrm{min}\sim 1/L$), and we get additional contributions to this sum, resulting in the finite size effect at large $R$.  That this effect is more noticeable at the higher stress $\tilde p$ (compare Fig.~\ref{R2varphi-def-I}b with \ref{R2varphi-def-I}a) is a consequence of the fact that the increase in $\chi(\mathbf{q})$ at small $|\mathbf{q}|$ becomes steeper at larger $\tilde p$ (see Fig.~\ref{chiI-vs-q-FS}).  If $\chi(\mathbf{q})$ eventually saturates to a constant as $|\mathbf{q}|\to 0$, these additional contributions as $N$ increases will become a negligible part of the sum, and the finite-size-effect will similarly become negligible.

\begin{figure}[h]
\begin{center}
\includegraphics[width=3.in]{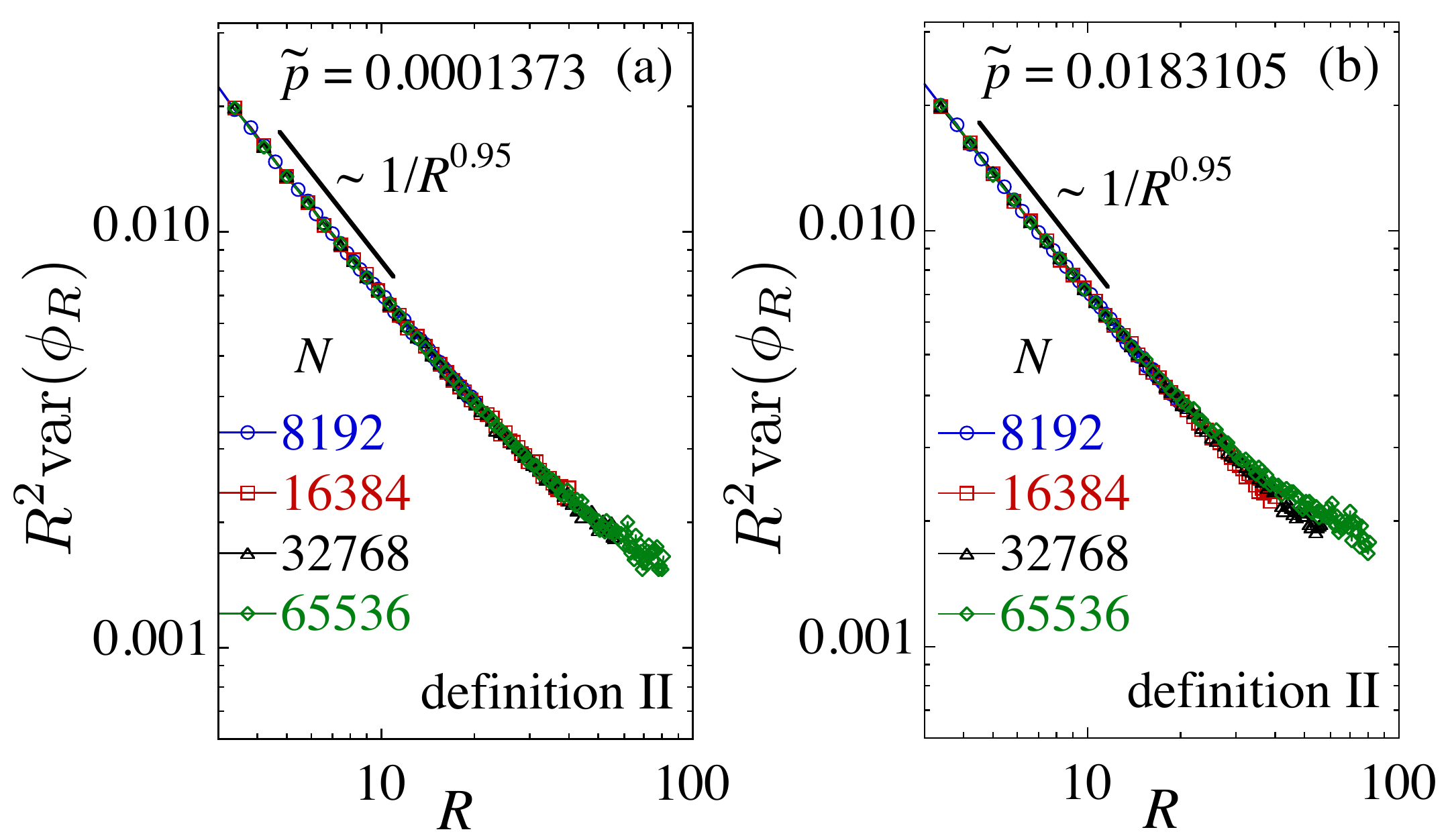}
\caption{(color online) $R^2\mathrm{var}(\phi_R)$ vs window radius $R$ for systems with $N=8192$ to $65536$ particles, at stress (a) $\tilde p=0.0001373$ and (b) $\tilde p=0.0183105$.  The power law decay at small $R$ is indicated.  Definition II of Eq.~(\ref{edefII}) for the local packing fraction $\phi(\mathbf{r})$ is used.  
}
\label{R2varphi-def-II}
\end{center}
\end{figure}

In Fig.~\ref{R2varphi-def-II} we similarly plot $R^2\mathrm{var}(\phi_R)$ vs $R$, but now using definition II of Eq.~(\ref{edefII}) for the local packing fraction.  Again we show results for several different system sizes, $N=8192$ to $65536$ for our smallest stress $\tilde p=0.0001373$, and largest stress $\tilde p= 0.0183105$.  The results are dramatically different from what is seen in Fig.~\ref{R2varphi-def-I}.  Here we see a much weaker dependence on the stress $\tilde p$, only a small finite size effect at the largest $R$, and a clear $R^{-1}$ decay over much of the range of data (a power-law fit to the data gives more precisely $\sim R^{-0.95}$).  Thus, while definition I gives no suggestion of hyperuniform behavior, definition II looks convincingly hyperuniform out to relatively large length scales $R$.  The dramatic difference in $\mathrm{var}(\phi_R)$ between the two definitions of the local packing fraction $\phi(\mathbf{r})$ is quite puzzling given the complete agreement of the corresponding $\chi(\mathbf{q})$ for the two definitions at small $|\mathbf{q}|$, as seen in Fig.~\ref{chi-1-2-B-vs-q}.  We can explain the reason for this difference in behavior as follows.

\begin{figure}[h]
\begin{center}
\includegraphics[width=3.4in]{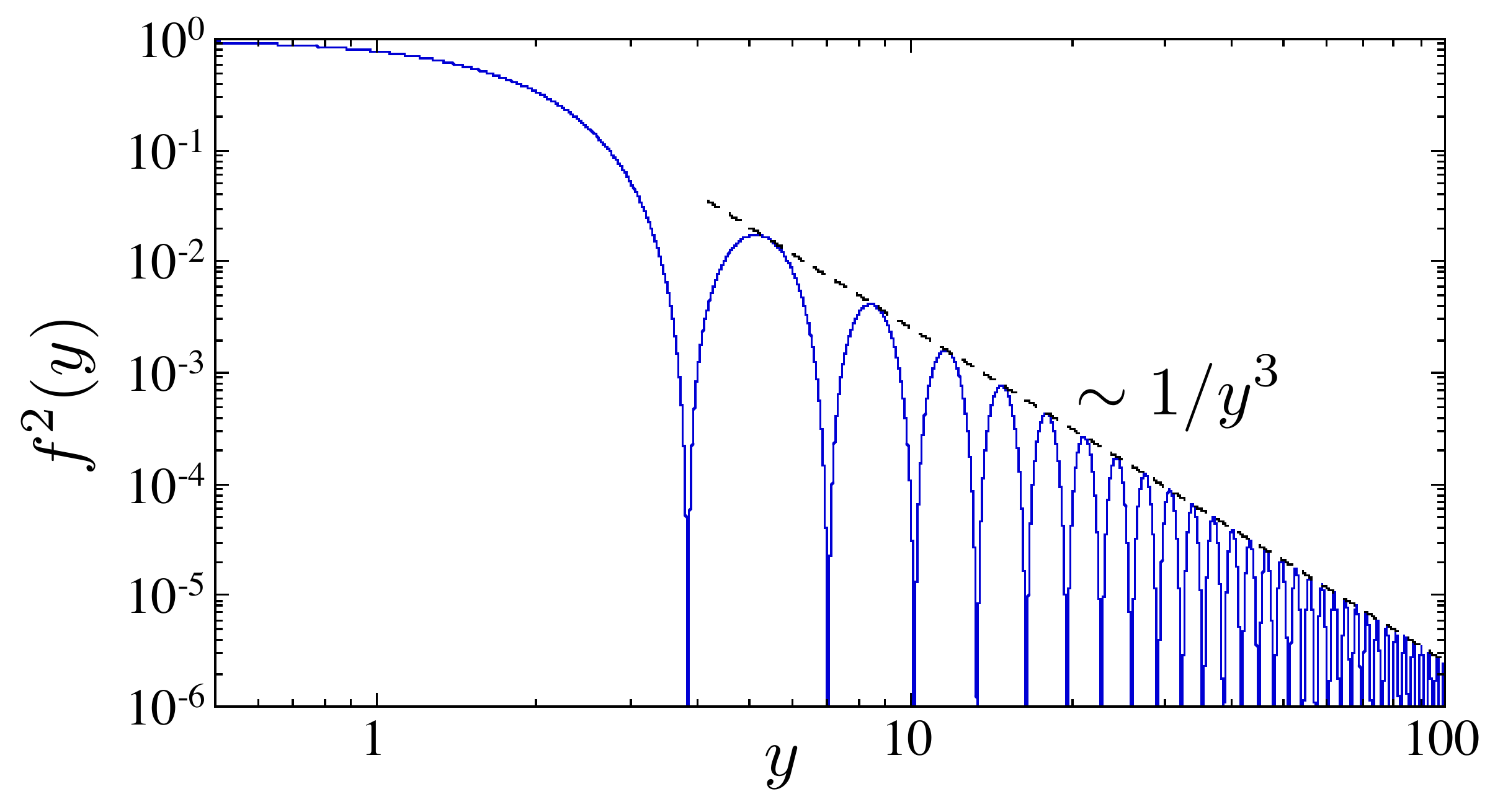}
\caption{(color online) Plot of $f^2(y)$ vs $y$, with the function $f(y)$ as defined in Eq.~(\ref{efy}).  The envelope of the oscillations at large $y$ decays as $1/y^3$.
}
\label{f2-vs-y}
\end{center}
\end{figure}

From Eqs.~(\ref{efy}) and (\ref{varphi-chi}) we can write the relation between $\mathrm{var}(\phi_R)$ and $\chi(\mathbf{q})$ as,
\begin{equation}
\mathrm{var}(\phi_R)=\frac{1}{V}\sum_{\mathbf{q}\ne 0}\chi(\mathbf{q})f^2(|\mathbf{q}|R),
\end{equation}
with $f(y)$ as defined in Eq.~(\ref{efy}).  Assuming that $\chi(\mathbf{q})$ depends only on $|\mathbf{q}|$ due to the average isotropy of the system \cite{qdirection}, we can integrate over the direction of $\mathbf{q}$ to get for our two dimensional system,
\begin{equation}
\mathrm{var}(\phi_R)=\frac{1}{L_y}\sum_{q\ne 0}\chi(q\mathbf{\hat y}) qf^2(qR)
\label{varphi-chi-2}
\end{equation}
with $q=2\pi n/L_y$.  In Fig.~\ref{f2-vs-y} we plot $f^2(y)$ vs $y$ on a log-log scale.  We see that for large $y$, it oscillates with a $1/y^3$ envelope. For an infinite system, if $\chi(\mathbf{q}\to 0)$ is a finite constant, then at sufficiently large $R$ a dimensional analysis  implies that $\mathrm{var}(\phi_R)$ {\em must} scale as $1/R^2$.  However for finite $R$, and in finite systems where the sum on $\mathbf{q}$ is discrete, the behavior of $\mathrm{var}(\phi_R)$ can depend in detail on the behavior of $\chi(\mathbf{q})$ at large $\mathbf{q}$; if the 
sum in Eq.~(\ref{varphi-chi-2}) is dominated by the large $\mathbf{q}$ terms, then we may find $\mathrm{var}(\phi_R)\sim 1/R^3$ because of the $1/(qR)^3$ dependence of $f^2(qR)$ and not because of any hyperuniformity of the system.

\begin{figure}[h]
\begin{center}
\includegraphics[width=3.4in]{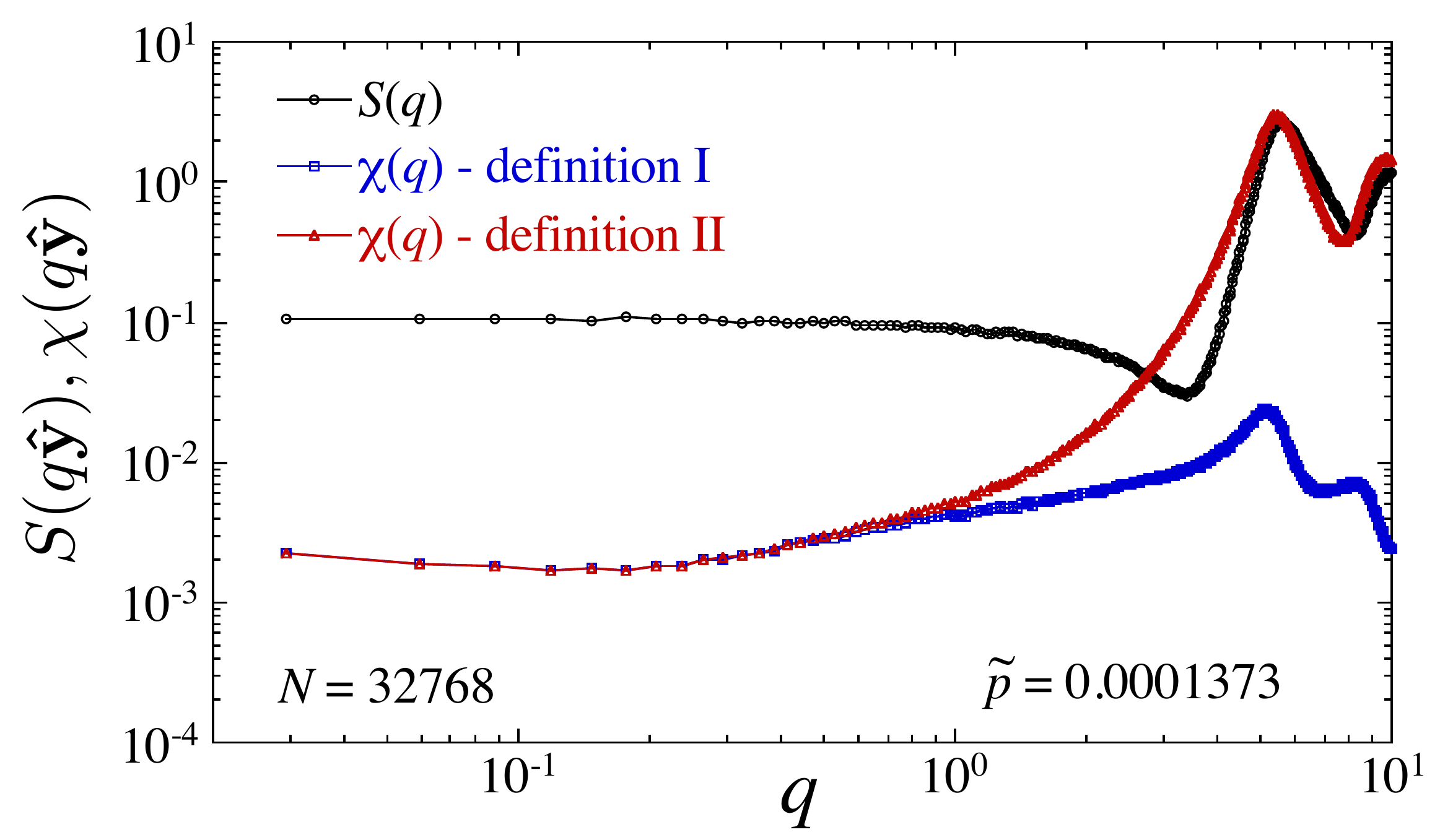}
\caption{(color online) Packing fraction fluctuation $\chi(q\mathbf{\hat y})$ as computed using definition I of Eq.~(\ref{edefI}), and definition II of Eq.~(\ref{edefII}), as well as the structure function $S(q\mathbf{\hat y})$, for a wide range of wavevectors up to $q=10$.  Results are shown for a system of $N=32768$ particles at our lowest stress $\tilde p=0.0001373$.  
}
\label{chiI-chiII-S-vs-bigq}
\end{center}
\end{figure}

The behavior of $\mathrm{var}(\phi_R)$ on observed length scales $R$ can thus be determined by the behavior of $\chi(\mathbf{q})$ at large wavevectors $\mathbf{q}$ with $|\mathbf{q}|>\pi/R$.  In Fig.~\ref{chiI-chiII-S-vs-bigq} we plot $\chi(q\mathbf{\hat y})$ for both definition I and definition II, as well as the structure function $S(q\mathbf{\hat y})$, for a much wider range of wavevectors, $0<q< 10$, than in previous plots.  We show results for $N=32768$ at our lowest stress $\tilde p = 0.0001373$.  As before, we see that $\chi(q\mathbf{\hat y})$ for the two definitions agree perfectly at small $q$, but then separate when $q\gtrsim 1$.  Moreover, $\chi(q\mathbf{\hat y})$ for definition II becomes roughly equal to $S(q\mathbf{\hat y})$, and over two orders of magnitude larger than that for definition I, when $q\gtrsim 5$.  Thus the contribution to $\mathrm{var}(\phi_R)$ from $\chi(q\mathbf{\hat y})$ at large $q$ should be expected to be more significant for definition II as compared to definition I.  

\begin{figure}[h]
\begin{center}
\includegraphics[width=3.4in]{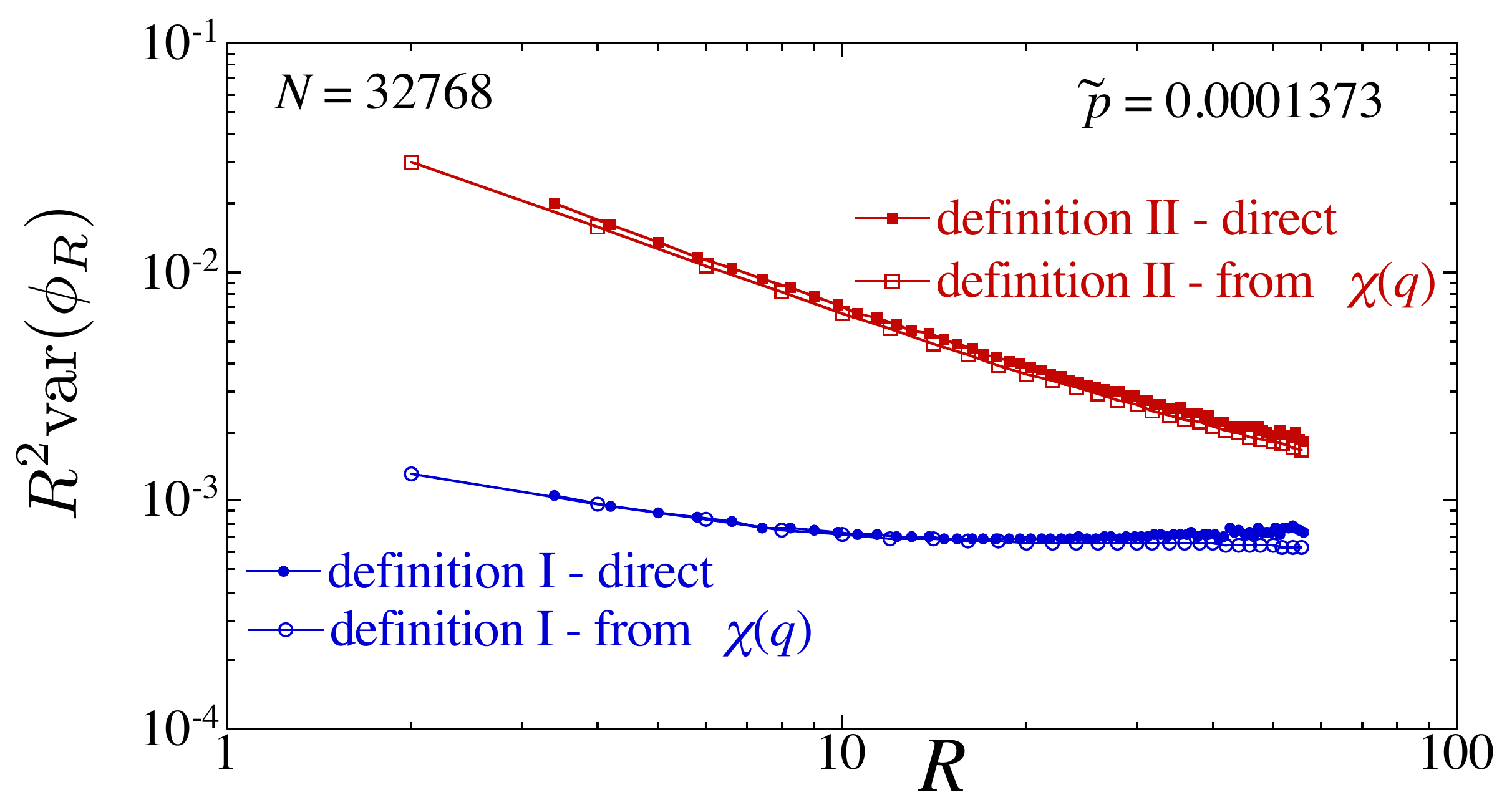}
\caption{(color online) $R^2\mathrm{var}(\phi_R)$ vs $R$, comparing the data of Figs.~\ref{R2varphi-def-I}a and \ref{R2varphi-def-II}a (direct computation) with the result from summing $\chi(q\mathbf{\hat y})$ over $q$ in Eq.~(\ref{varphi-chi-2}), for both definition I and definition II of the local packing fraction $\phi(\mathbf{r})$.  Results are for a system with $N=32768$ particles at our smallest stress $\tilde p=0.0001373$.
}
\label{varphi-chiq-compare}
\end{center}
\end{figure}

To check this, we compute $\mathrm{var}(\phi_R)$ by explicitly summing the series in Eq.~(\ref{varphi-chi-2}), using the data for $\chi(q\mathbf{\hat y})$ from Fig.~\ref{chiI-chiII-S-vs-bigq}.  Our results for $R^2\mathrm{var}(\phi_R)$ vs $R$ are shown in Fig.~\ref{varphi-chiq-compare} for $N=32768$ at $\tilde p=0.0001373$. We compare these results to the direct computation of $\mathrm{var}(\phi_R)$ as shown in Figs.~\ref{R2varphi-def-I}a and \ref{R2varphi-def-II}a.  For definition I we find excellent agreement between the series and the direct computation when we sum the series  up to $q=10$.  For definition II we find that we must sum even more terms, up to $q=50$, in order to get reasonable agreement.  We thus see again that the large $q$ (i.e. small length scale) fluctuations are larger, and so contribute more to $\mathrm{var}(\phi_R)$, for definition II than for definition I.
Our results in Fig.~\ref{varphi-chiq-compare} show that the computation of $\mathrm{var}(\phi_R)$ in Fig.~\ref{varphi-def-I-def-II} is indeed consistent with our computation of $\chi(q\mathbf{\hat y})$ in Fig.~\ref{chi-1-2-B-vs-q}, and that the reason for the dramatic difference in $\mathrm{var}(\phi_R)$, comparing definition I with definition II, is the influence of fluctuations at moderately large $\mathbf{q}$, which persist even to large $R$.  We conclude that $\chi(\mathbf{q})$, rather than $\mathrm{var}(\phi_R)$, is the better measure to use to check for hyperuniformity in our two dimensional system.

\section{Discussion and conclusions}
\label{secDiscuss}

In this work we have considered the fluctuations of density and packing fraction in mechanically stable packings of bidisperse frictionless particles at finite stress.  A distinguishing feature of our work is that we simulate at fixed {\em isotropic} global stress, rather than at fixed packing fraction.  We investigate states 
{\em above} the jamming transition,  in contrast to earlier works \cite{Berthier, Zachary2, Zachary3}  that considered the case of packings exactly at the jamming $\phi_J$.  

\subsection{Comparison to previous works}

Berthier et al. \cite{Berthier} considered both an experimental two dimensional system of $N=8000$ bidisperse particles, and a numerical three dimensional system of $N=64000$ soft-core particles of varying size dispersities.  This corresponds to system lengths of roughly 90 and 40 particle diameters in the experimental and numerical systems respectively.  The experimental system was jammed under slow compression.  The numerical configurations were created  starting from jammed states above $\phi_J$, and then slowly decompressing until the system unjammed, followed by a slower recompression until the system jammed again, as measured by a finite energy per particle of order $10^{-11}$.  In both experimental and numerical systems, a $\chi_T(q)$ is observed that appears to linearly decrease towards zero.  However, in both cases the measured $\chi_T(q)$ only extends down to roughly $qd\approx 0.15$, where $d$ is the average diameter of the particles, and the data is quite scattered below $qd\approx 0.5$.  These results thus give evidence for hyperuniformity out to length scales $\ell/d\sim 2\pi/0.5 \approx 12$, but not necessarily on longer length scales.  

Zachary et al. \cite{Zachary2, Zachary3} considered a much larger numerical system in two dimensions, with up to $N=10^6$ particles (and so system length of roughly 1000 particle diameters).  They used a bidisperse system with particles of different shapes, with size ratio 1.4 and a concentration of small particles $x_s=0.75$, and large particles $x_b=0.25$.  They used the Lubachevsky-Stillinger algorithm \cite{Lubachevsky} to generate their jammed states.  This is an event-driven molecular dynamics for elastic hard-core particles, where particles are  inflated at a prescribed rate from an initial thermally equilibrated dilute state so as to rapidly quench the hard-core gas into a thermal glassy state.  The particle inflation continues until the system seizes up into a strictly jammed state 
that they denote as {\em maximally random jammed} (MRJ).  They measure both $\chi(q)$ and $\mathrm{var}(\phi_R)$ (using definition I) in the MRJ state, and for circular particles they find strong evidence from $\mathrm{var}(\phi_R)$ for hyperuniformity out to length scales $2R/d\sim 80$.

More recently, other works have reconsidered hyperuniformity in systems of {\em monodisperse} spheres in three dimensions, and have considered behavior approaching, rather than strictly at, $\phi_J$.  Recall, for monodisperse systems, hyperuniformity is indicated by the behavior of the structure function, $S(\mathbf{q})\sim |\mathbf{q}|$ as $|\mathbf{q}|\to 0$.  Hopkins et al. \cite{Hopkins}, using the same protocol as Zachary et al. \cite{Zachary2,Zachary3} for a system with $N=10^6$ particles (system length roughly 100 particle diameters), measure $S(\mathbf{q})$ at various $\phi$ approaching $\phi_J$ from below.  They find $S(\mathbf{q})\approx aq+b$, with $b\to 0$ as $\phi\to\phi_J$, and from this extract a length scale $\xi\sim 1/b^{1/3}$ that diverges as jamming is approached and the system becomes hyperuniform.


Ikeda and Berthier \cite{Ikeda} study $N=512000$ monodisperse soft-core particles in three dimensions.  Starting from a random configuration of particles in a fixed cubic box at packing fraction $\phi=0.8$, well above $\phi_J\approx 0.646$, they use the FIRE algorithm \cite{FIRE} to minimize the interaction energy and obtain a mechanically stable state.  They then decrease the particle density in small steps, energy minimizing at each step, to obtain configurations spanning a range of packing fractions from $\phi=0.8$ to just above the jamming $\phi_J$.  Their results are averaged over 8 independent starting configurations.  Computing $S(\mathbf{q})$ they find, for all but their largest value of $\phi=0.8$, that data at the different $\phi$ essentially overlap and are linear in $q$, as expected for a hyperuniform system, over an extended range of $0.4<q<7$.  However at their smallest $q$, they find $S(\mathbf{q})$ saturates to a finite value $\sim 10^{-3}$, similar in magnitude to what we have found in the present work for $\chi(\mathbf{q})$; they, however, see a plateau in $S(\mathbf{q})$ at small $q$ rather than the minimum that we find in $\chi(\mathbf{q})$.  Ikeda and Berthier thus conclude that hyperuniformity is only weakly dependent on packing fraction $\phi$, but persists out to only a finite length scale $\approx 15d$.  Ikeda and Berthier further find that this behavior is stable to the addition of small finite thermal fluctuations.

The above results, combined with our own, suggest that mechanically stable jammed packings {\em above} $\phi_J$ do {\em not} display hyperuniform fluctuations  of the packing fraction out to arbitrarily large length scales, but are hyperuniform only out to a finite $\ell^*(\phi)$ that is weakly dependent on $\phi$ and does {\em not} appear to diverge as $\phi\to\phi_J$ from above. However the results of Zachary et al. \cite{Zachary2,Zachary3} and Hopkins et al. \cite{Hopkins} suggest that hyperuniformity may exist in hard-core particle systems, when compressed to $\phi_J$ from {\em below}.  It may therefore be that the presence or absence of hyperuniformity out to arbitrarily large length scales depends on the specific protocol used to construct the jammed state at $\phi_J$.  We also cannot rule out the possibility that hyperuniformity may still exist in jammed packings above $\phi_J$, but restricted to a region closer to $\phi_J$ than we have been able to explore in this work.


\subsection{Alternative ensembles}

\begin{figure}
\includegraphics[width=3.4in]{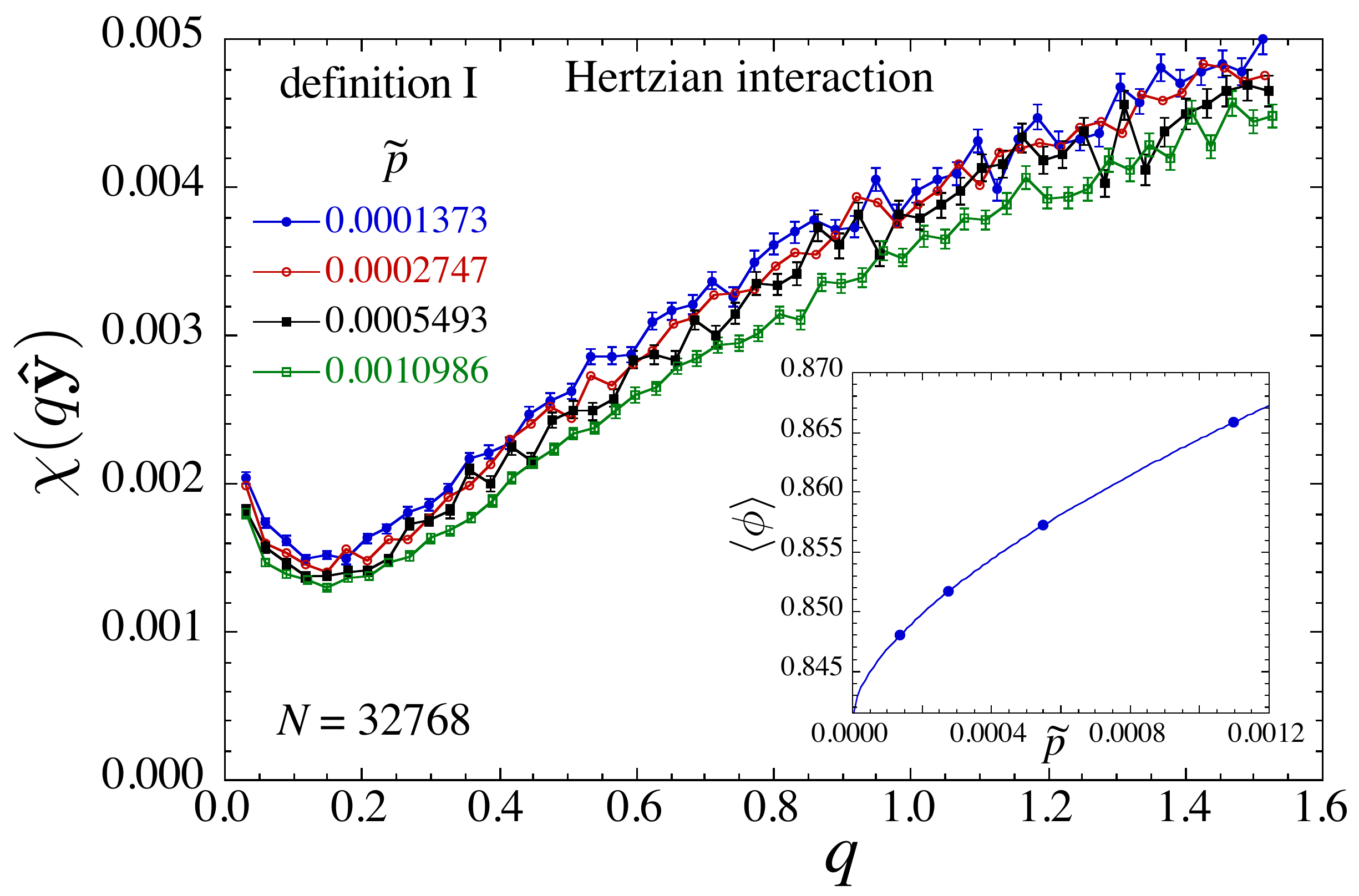}
\caption{(color online) Fluctuation in packing fraction $\chi(q\mathrm{\hat y})$ vs $q$, using the definition I of Eq.~(\ref{edefI}) for the local packing fraction $\phi(\mathbf{r})$, for the case of a Hertzian interaction ($\alpha=5/2$ in Eq.~(\ref{einteraction})).  Results are shown for several different values of the stress per particle $\tilde p$, for a system with $N=32768$ particles.  The inset shows the average packing fraction $\langle\phi\rangle$ as a function of $\tilde p$.
}
\label{chi-Hertz}
\end{figure}

To check how sensitive our results for $\phi>\phi_J$ are to the particular system we have used above, we have considered two other ensembles.  The first is to use a Hertzian interaction, with $\alpha=5/2$ in Eq.~(\ref{einteraction}), in place of the harmonic interaction.  All other details of the system remain the same.  
In Fig.~\ref{chi-Hertz} we plot $\chi(q\mathbf{\hat y})$, computed according to definition I of Eq.~(\ref{edefI}), vs $q$ for the four lowest $\tilde p$ that we used for the harmonic interaction.  We use a system size with $N=32768$ particles.  For the Hertzian interaction, pressure is expected \cite{OHern} to scale with packing fraction according to $\tilde p\sim (\phi-\phi_J)^{3/2}$, hence the $\langle \phi\rangle$ for the Hertzian system (shown as the inset to Fig.~\ref{chi-Hertz}) is larger than that of the harmonic system at equal values of $\tilde p$.  The Hertzian interaction further differs from the hamonic in that for the Hertzian system the bulk modulus vanishes continuously as $\phi\to\phi_J$ from above, while for the harmonic system the bulk modulus approaches a finite constant as $\phi\to\phi_J$ from above, and then jumps discontinuously to zero below $\phi_J$ \cite{OHern}.  Nevertheless, we find that $\chi(q\mathbf{\hat y})$ for the Hertzian system is qualitatively the same as for the harmonic case, with a well defined minimum that does not appear to be moving to smaller $q$ as $\tilde p$ decreases.

\begin{figure}
\includegraphics[width=3.4in]{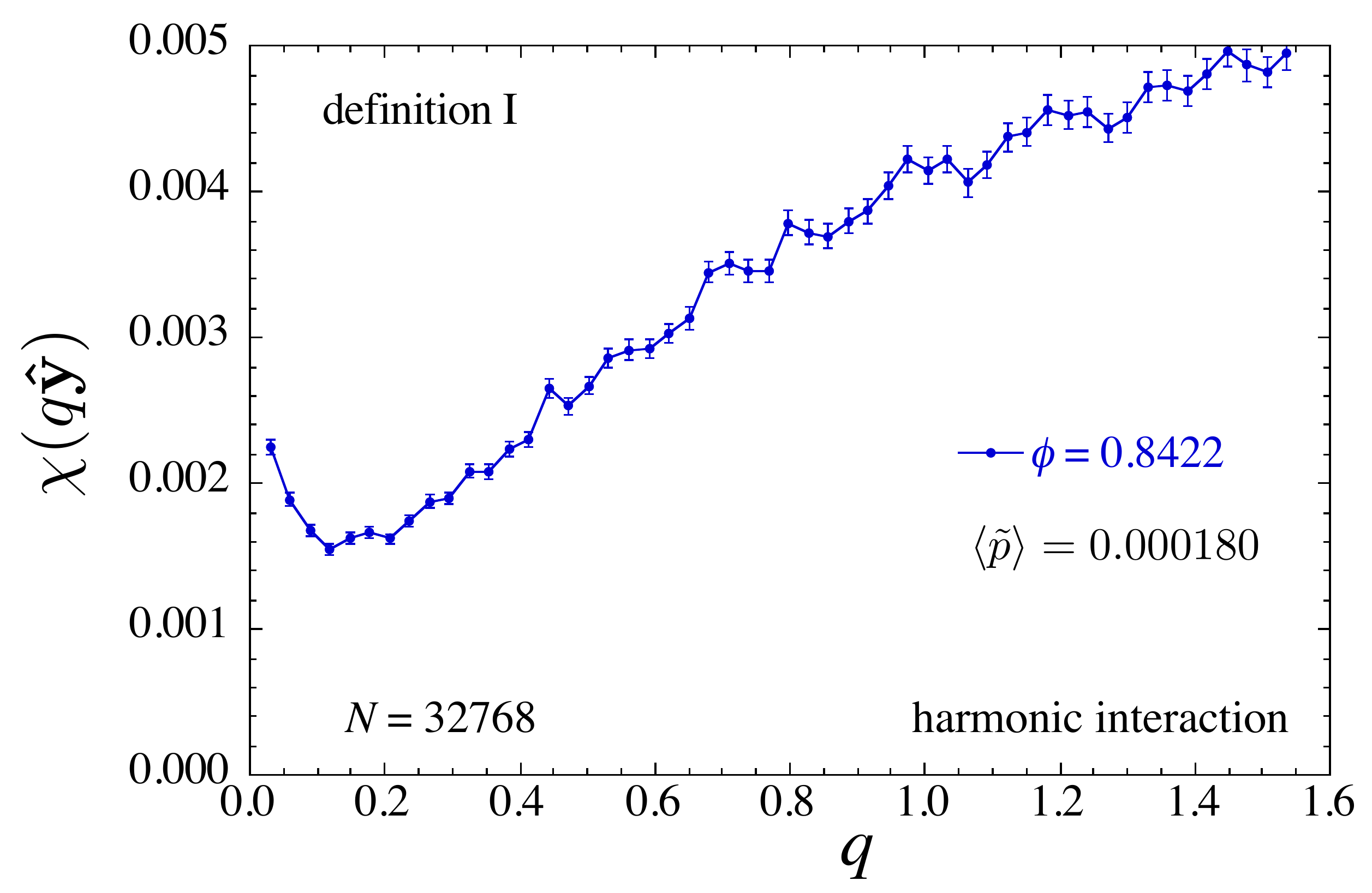}
\caption{(color online) Fluctuation in packing fraction $\chi(q\mathrm{\hat y})$ vs $q$, using the definition I of Eq.~(\ref{edefI}) for the local packing fraction $\phi(\mathbf{r})$, for the case of the harmonic interaction in an ensemble at fixed global packing fraction $\phi=0.8422$. 
Results are shown for a system with $N=32768$ particles.  
}
\label{chi-fixed_phi}
\end{figure}

The second is to consider the harmonic interaction, but to obtain our configurations by quenching at fixed $\phi$ within a fixed square box.  Such constant $\phi$ ensembles have usually been used in earlier works \cite{Berthier,Zachary2,Zachary3,Hopkins, Ikeda}.  Unlike the constant stress ensemble, where configurations all have the same $\tilde p$ and so can be viewed as all at the same distance from the jamming transition $\tilde p=0$, the constant $\phi$ ensemble has a fluctuating $\tilde p$ and so different configurations $i$ are at  different distances from their configuration specific jamming transition $\phi_{Ji}$ \cite{OHern,Pinaki}.  The constant $\phi$ ensemble in a fixed box also allows there to be a finite residual shear stress in the quenched configuration \cite{Dagois}.  In Fig.~\ref{chi-fixed_phi} we plot $\chi(q\mathbf{\hat y})$, computed according to definition I of Eq.~(\ref{edefI}), vs $q$ for the case $\phi=0.8422$ close to $\phi_J\approx 0.84159$.  We use a system size with $N=32768$ particles.  Again we see qualitatively the same behavior as before.

Note, the fixed value of $\phi=0.8422$ in Fig.~\ref{chi-fixed_phi} was chosen as it is equal to the $\langle \phi\rangle$ for a system with $\tilde p=0.0002747$ in the fixed stress ensemble (our next to lowest value of $\tilde p$).  However in the fixed $\phi=0.8422$ ensemble, we find that the average stress per particle is $\langle\tilde p\rangle=0.000180$, lower than the corresponding value in the fixed stress ensemble.  
This suggests that the jamming density $\phi_J$ of the constant $\phi$ ensemble is slightly larger than the jamming density of the constant stress ensemble.  That is consistent with our estimate of $\phi_J\approx 0.84159$ for the constant stress ensemble from Fig.~\ref{phi-vs-p}, as compared with the estimate of $\phi_J\approx 0.84177$ for the constant $\phi$ ensemble from Ref.~\cite{VagbergFSS}.
We also note that the width of the distribution of $\tilde p$ found in this fixed $\phi$ ensemble is  rather large, $\sqrt{\langle\tilde p^2\rangle-\langle\tilde p\rangle^2}/\tilde p =0.40$, while the corresponding width of the distribution of the residual deviatoric stress per particle $\tilde\sigma$ is rather small, $\sqrt{\langle\tilde\sigma^2\rangle}/\tilde p=0.00053$.

\subsection{Rattlers and polydispersity}

\begin{figure}[h]
\begin{center}
\includegraphics[width=3.4in]{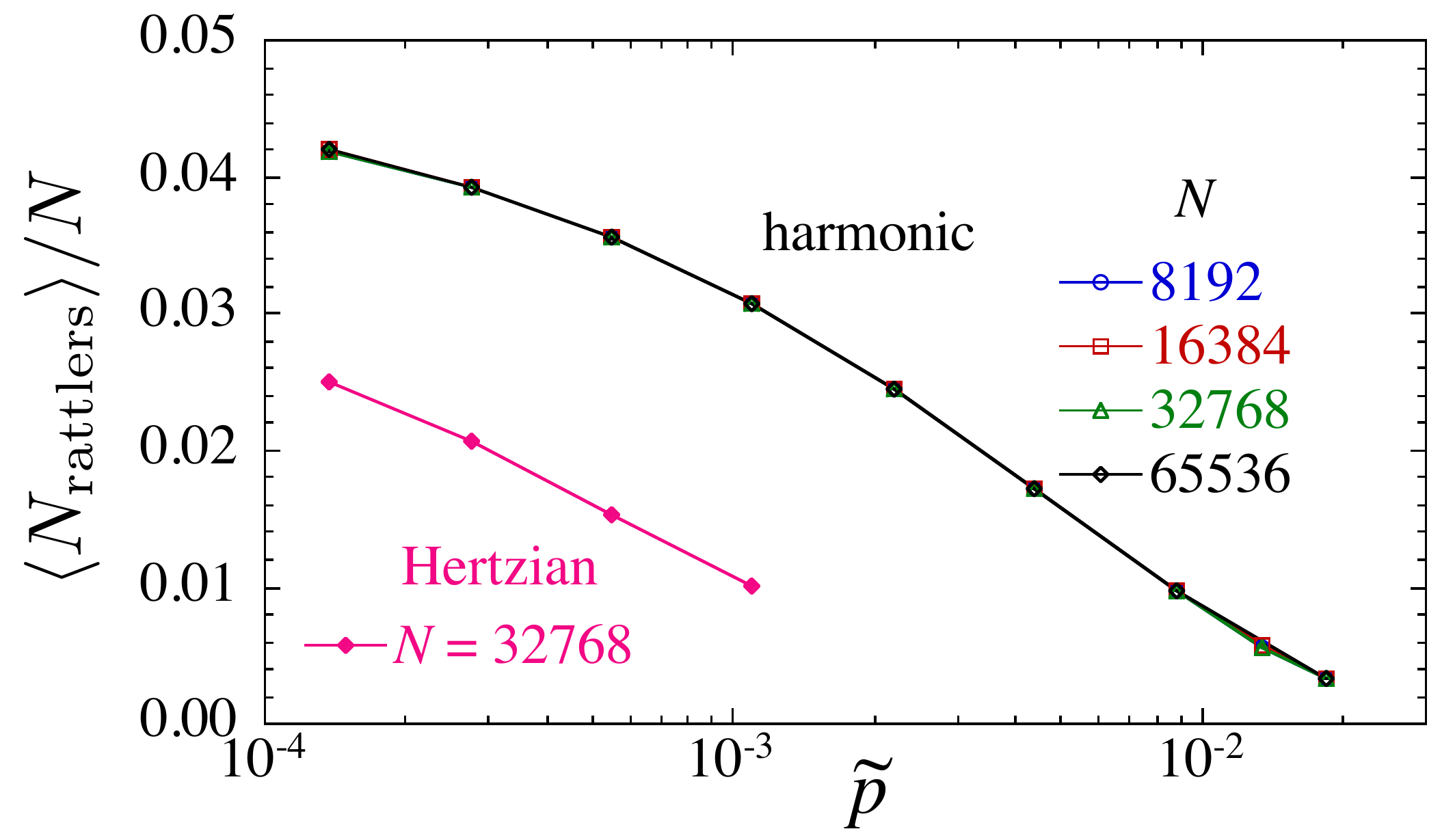}
\caption{(color online) Fraction of the particles that are rattlers, $\langle N_\mathrm{rattlers}\rangle/N$ vs stress per particle $\tilde p=\Gamma_N/N$.  For the harmonic interaction we show results for systems with $N=8192$ to $65536$ particles; the points for different $N$ in the figure overlap each other.  For the Hertzian interaction we show results for $N=32768$.
}
\label{N_ratters-o-N}
\end{center}
\end{figure}

It has been suggested \cite{Donev1,Ikeda} that rattlers may play a role in the breaking of hyperuniformity on large length scales.  Rattlers result when a particle has an insufficient number of contacts to constrain its motion in all directions.  Determining the number of rattlers according to the method described in Sec.~\ref{sGlobal},
in Fig.~\ref{N_ratters-o-N} we plot the fraction of particles that are rattlers $\langle N_\mathrm{rattlers}\rangle/N$ vs the stress per particle $\tilde p$.  For the harmonic interaction, we plot results for  systems with $N=8192$ to $65536$ particles.  For the Hertzian interaction, we plot results for $N=32768$ only.  We see that $\langle N_\mathrm{rattlers}\rangle/N$ is independent of the system size $N$, and {\em decreases} with increasing $\tilde p$.  For the harmonic interaction, $\langle N_\mathrm{rattlers}\rangle/N$ changes by an order of magnitude over the range of $\tilde p$ we study.  If rattlers were responsible for the breaking of hyperuniformity, we might expect that the length $\ell^*$ to which hyperuniformity extends should increase as the density of rattlers decreases, i.e. as $\tilde p$ increases.  However our results in Fig.~\ref{chiI-vs-q} show exactly the opposite trend; the $q^*$ that locates the minimum of $\chi(q\mathbf{\hat y})$ increases slightly with increasing $\tilde p$, and so $\ell^*=2\pi/q^*$ {\em decreases} with increasing $\tilde p$.  Our results thus provide no obvious relation between rattlers and the breaking of hyperuniformity.

Another possibility that might lead to the breaking of hyperuniformity is suggested \cite{referee} by the work of Dreyfus et al. \cite{Dreyfus}.  Their work is primarily concerned with the detection of hyperuniformity in experimental systems, where particles are polydisperse, and the exact size of individual particles is not a priori known but must be determined by optical measurements.  Errors in the determination of the exact particle sizes were found to result in an apparent breaking of hyperuniformity at small wavevectors (large length scales).  As an extreme example of this effect, one can consider the error introduced if, in a bidisperse or polydisperse system, one approximated all particles as having the same average size.  In that approximation, the packing fraction fluctuation $\chi(\mathbf{q})$ just becomes proportional to the structure function $S(\mathbf{q})$, which clearly does not show hyperuniformity at small $q$, as seen in Fig.~\ref{S-vs-q}.

\begin{figure}[h]
\begin{center}
\includegraphics[width=3.4in]{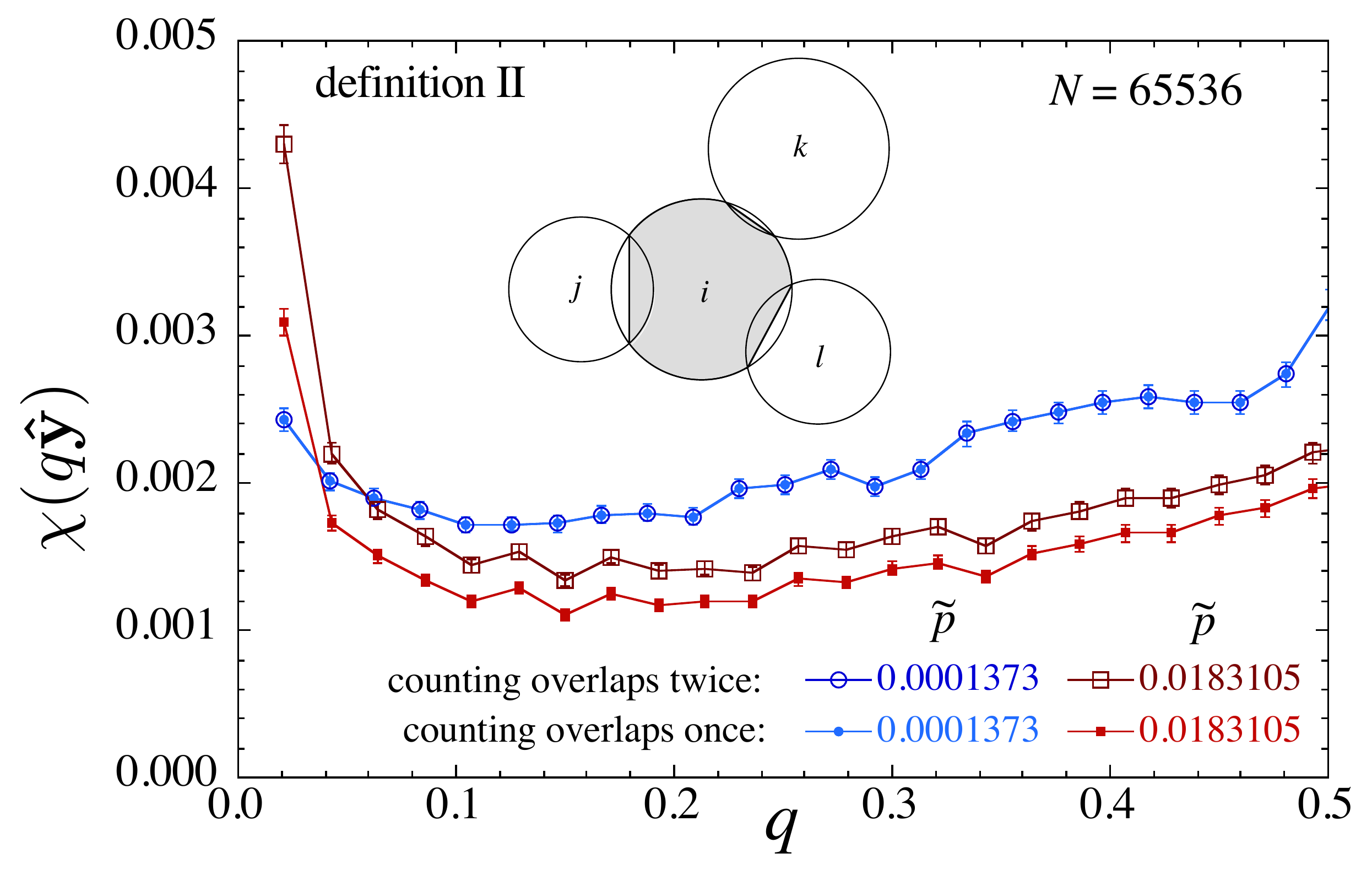}
\caption{(color online) Packing fraction fluctuation $\chi(q\mathbf{\hat y})$ vs $q$, using definition II of Eq.~(\ref{edefII}) for a system of $N=65536$ particles, at two different values of the stress per particle $\tilde p$.  We compare results where the weight of each particle is taken as the circular area of the isolated particle (denoted as ``counting overlaps twice"), vs where the weight of each particle is taken as the non-overlapping part of that circular area, as illustrated by the shaded region for particle $i$ in the inset (denoted as ``counting overlaps once").  A small difference is seen between these two sets of weights at the larger value of $\tilde p$, but not at the smaller value.
}
\label{overlapCompare65}
\end{center}
\end{figure}

In our bidisperse particle simulations, we of course know the position and size of each and every particle exactly.  Nevertheless, an effective polydispersity may be viewed to arise from the following effect.  Both our definitions I and II count each particle with a weight equal to the area of the particle in isolation. However in our jammed packings, particles in contact necessarily have some amount of overlap.  Our definitions I and II therefore count this overlap area twice, once for each particle.   One might imagine that a more ``correct" definition of the local packing fraction should count this overlap area only once, dividing it proportionally between the two contacting particles.  For example, as sketched in the inset to Fig.~\ref{overlapCompare65}, particle $i$ should have a weight equal to only the shaded area, rather than the full area of the corresponding circle.  If $v_i=\pi (d_i/2)^2$ is the area of the circle of particle $i$, then the weight with which particle $i$ enters the local packing fraction should instead be taken as $\tilde v_i\equiv v_i-\sum_j^\prime \delta v_{ij}$, with $\delta v_{ij}$ the area subtracted due to the overlap with particle $j$.  The weights $\tilde v_i$ are therefore polydisperse, depending on the varying overlaps in the system.  If one computes $\chi(\mathbf{q})$ using the bidisperse weights $v_i$ rather than the more correct polydisperse weights $\tilde v_i$, it could lead to a breaking of hyperuniformity that is only apparent, i.e. a consequence of using incorrect weights.

However, if $\delta_{ij}\equiv (d_i+d_j)/2-r_{ij}$ is the overlap length of the contact, then $\delta v_{ij}/v_i\propto  (\delta_{ij}/d_i)^{3/2}\propto p^{3/2}$, where the last result follows since the pressure $p\sim \langle\delta_{ij}\rangle$ for the harmonic interaction potential.  Thus this effect should vary with the pressure and vanish continuously as $p\to 0$, as one approaches the jamming transition.
To test this notion, we have therefore computed $\chi(q\mathbf{\hat y})$ according to definition II of Eq.~(\ref{edefII}), but using the weights $\tilde v_i$ as described above, computed exactly for each particle according to its own specific overlaps.  We use definition II since it is easier to implement than definition I, in the case where each particle has a unique, nonsymmetric (i.e. circle minus overlaps), shape.  However we expect from Fig.~\ref{chi-1-2-B-vs-q} that $\chi(q\mathbf{\hat y})$ will be identical for definitions I and II at the small $q$ of interest.  In Fig.~\ref{overlapCompare65} we plot the resulting $\chi(q\mathbf{\hat y})$ for our largest system with $N=65536$ particles, at both our smallest and largest values of $\tilde p$.  We compare the $\chi(q\mathbf{\hat y})$ obtained from using the original weights $v_i$ (denoted as ``counting overlaps twice") with that using the new weights $\tilde v_i$ (denoted as ``counting overlaps once").

At the largest $\tilde p=0.0183105$, we see a clear shift between the results from the different sets of weights, however the qualitative behavior remains the same, with a clear minimum at the same $q^*$, and $\chi$ increasing as $q$ decreases below $q^*$.  At our smallest $\tilde p=0.0001373$, however, the results from the two sets of weights are essentially equal.  Thus taking overlaps into account does not result in a restoration of hyperuniformity on large length scales, and the insensitivity of our results to the different choices of weights at our smallest $\tilde p$ is yet another indication that our smallest pressures are, by all relevant measures, quite close to jamming.

Finally, it is interesting to note that the experiments on PINIPAM microgel particles, reported on in Dreyfus et al. \cite{Dreyfus}, may actually correspond more closely to our conclusions than to the claim in favor of hyperuniformity.  
As these authors note, 
the small $q$ behavior of $\chi(q)$ for PINIPAM, shown in the inset to their Fig.~9b, does not suggest hyperuniformity; indeed it is qualitatively similar to what we see in our Fig. 7.  However the $q^*$ at which $\chi(q)$ has its minimum is so much larger in the experiments of Dreyfus et al. than what we find here, that in their case it may well be an artifact of system size, such as Dreyfus et al. claim.
However, if we consider the real space decay of $\mathrm{var}(\phi_R)$ with $R$, our results in Fig.~\ref{R2varphi-def-I} for definition I (corresponding to the usage in Dreyfus et al.) show that the initial decay, before the minimum is reached, is  $\mathrm{var}(\phi_R)\sim R^{-\lambda}$, with $\lambda\approx 2.4$ for our smaller $\tilde p$, and $\lambda\approx 2.3$ for our larger $\tilde p$.  This is not far from the value $\lambda\approx 2.2$ reported in Dreyfus et al. for a similar range of $R$, using their j-PSR reconstruction as shown in their Fig.~9a.  Yet in our case, our $\lambda>2$ does {\em not} demonstrate that the system is hyperuniform; we see hyperuniformity is broken only by looking at larger length scales.
This comparison thus suggests that the PINIPAM experiments may actually be above the jamming $\phi_J$, and are not inconsistent with the absence of hyperuniformity on long length scales.

\section*{Acknowledgments}

This work was supported by NSF Grant No. DMR-1205800.  Computations were carried out at the Center for Integrated Research Computing at the University of Rochester.  We wish to thank L.~Berthier, P.~Chaudhuri and A.~J.~Liu for helpful discussions.  We also thank L.~Berthier for sharing with us a preliminary version of Ref.~\cite{Ikeda}.

\section*{Appendix}

In this appendix we provide some further details about the minimization procedure of Sec.~\ref{SecModel} that we use to obtain our mechanically stable configurations at fixed isotropic stress.

\begin{figure}[h]
\begin{center}
\includegraphics[width=3.4in]{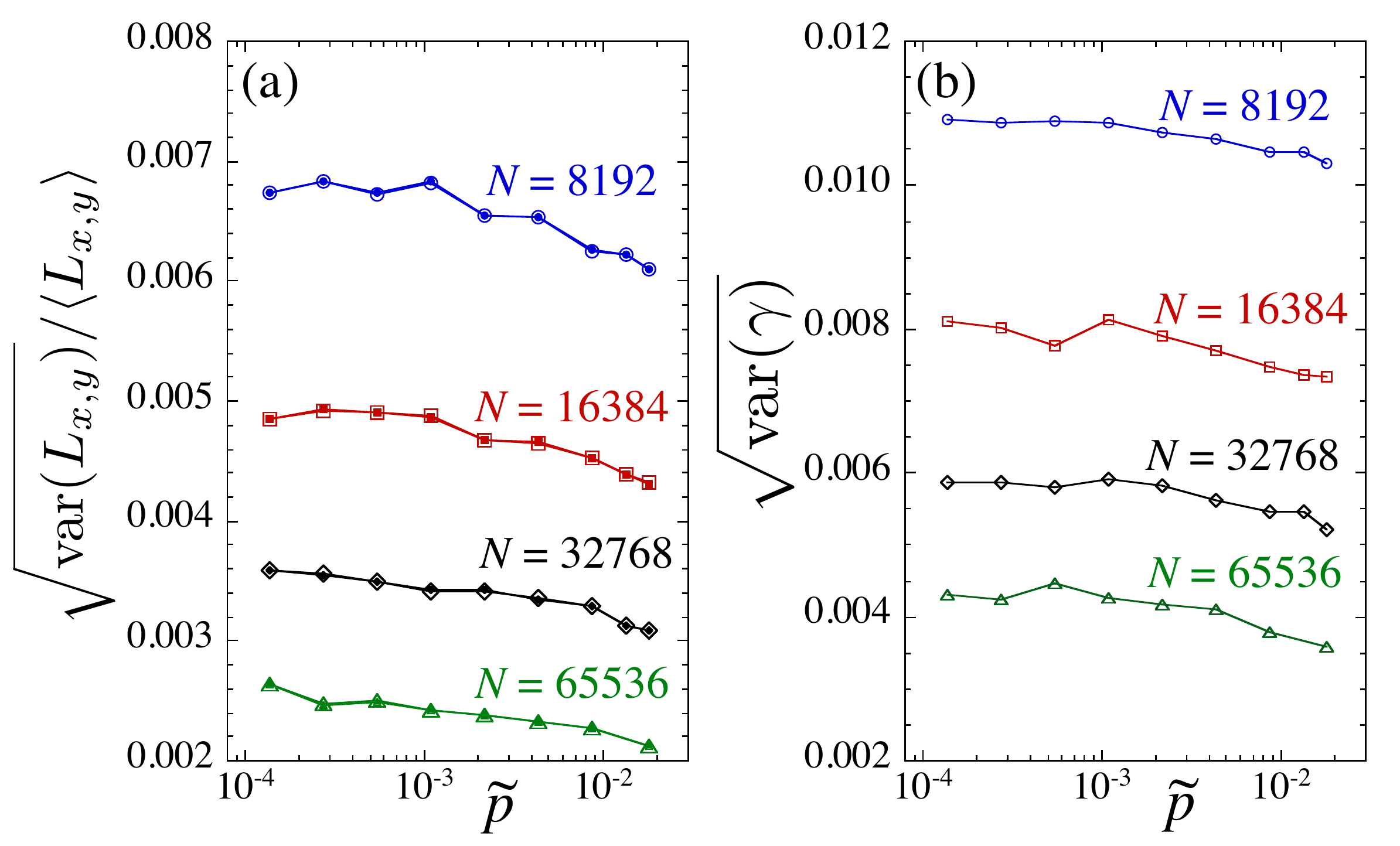}
\caption{(color online) (a) Relative fluctuations in the system box lengths $L_x$ and $L_y$ vs stress per particle $\tilde p$, for system sizes $N=8192$ to $65536$. Solid symbols show the fluctuations in $L_x$, while open symbols show $L_y$.  (b) Fluctuations in the dimensionless box skew parameter $\gamma$ vs $\tilde p$. See Fig.~\ref{f2} for the definition of parameters $L_x$, $L_y$, $\gamma$.  In both cases the fluctuations scale as $1/\sqrt{N}$.
}
\label{boxFlucs}
\end{center}
\end{figure}

\begin{figure}[h]
\begin{center}
\includegraphics[width=3.2in]{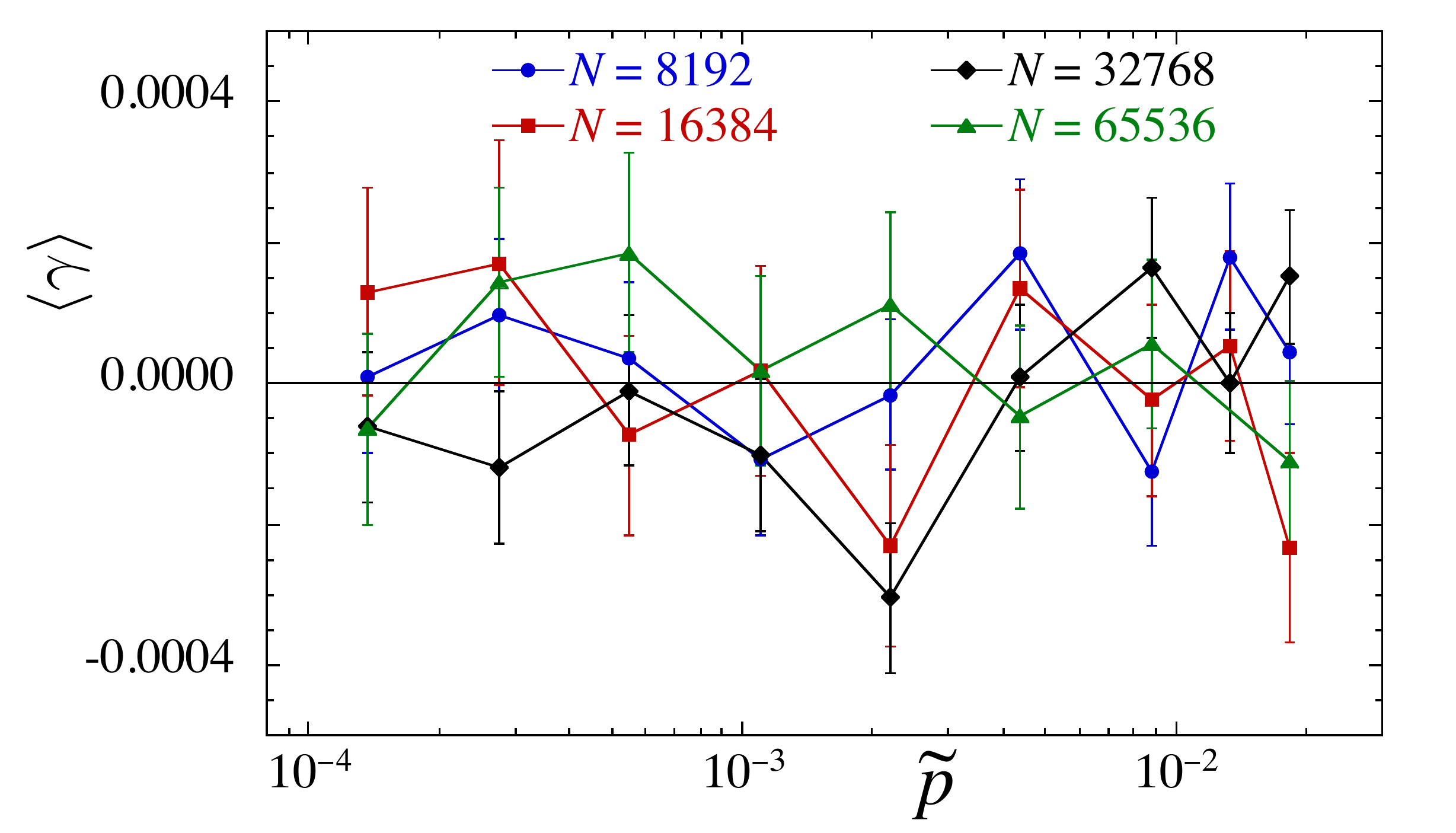}
\caption{(color online) Average box skew $\langle\gamma\rangle$ vs stress per particle $\tilde p$, for system sizes $N=8192$ to $65536$. Error bars represent one standard deviation of estimated statistical error, showing that $\langle\gamma\rangle=0$ within the estimated errors.
}
\label{gammaAve-vs-p}
\end{center}
\end{figure}

Since our minimization procedure is carried out at fixed total system stress $\Sigma_{\alpha\beta}$, the system box parameters $L_x$, $L_y$ and $\gamma$ (see Fig.~\ref{f2}) will vary from specific minimized configuration to configuration.  In Fig.~\ref{boxFlucs} we show the extent of these variations for the different system sizes $N=8192$, $16384$, $32768$ and $65536$.  In Fig.~\ref{boxFlucs}a we show the relative fluctuations in box lengths, $\sqrt{\mathrm{var}(L_x)}/\langle L_x\rangle$ and $\sqrt{\mathrm{var}(L_y)}/\langle L_y\rangle$ vs the stress per particle $\tilde p=\Gamma_N/N$.  Solid symbols are for $L_x$ while open symbols are for $L_y$.  Since the system is on average isotropic, we expect the fluctuations in $L_x$ and $L_y$ to be equal, and we indeed find that to be so.  The fluctuations are also found to scale as $1/\sqrt{N}$, as would naively be expected.  In Fig.~\ref{boxFlucs}b we show the fluctuations in the dimensionless skew parameter, $\sqrt{\mathrm{var}(\gamma)}$ vs $\tilde p$.  The size of the fluctuations in $\gamma$ are 
slightly larger but comparable to the fluctuations in the box lengths.  Again we find that the fluctuations scale as $1/\sqrt{N}$.
We also note that, as expected, the average skew $\langle\gamma\rangle=0$ within the estimated statistical error, as shown in Fig.~\ref{gammaAve-vs-p}.

Our minimization procedure necessarily produces the desired isotropic stress configurations only to a certain numerical accuracy.  We now provide details of the degree of that accuracy.

\begin{figure}[h]
\begin{center}
\includegraphics[width=3.4in]{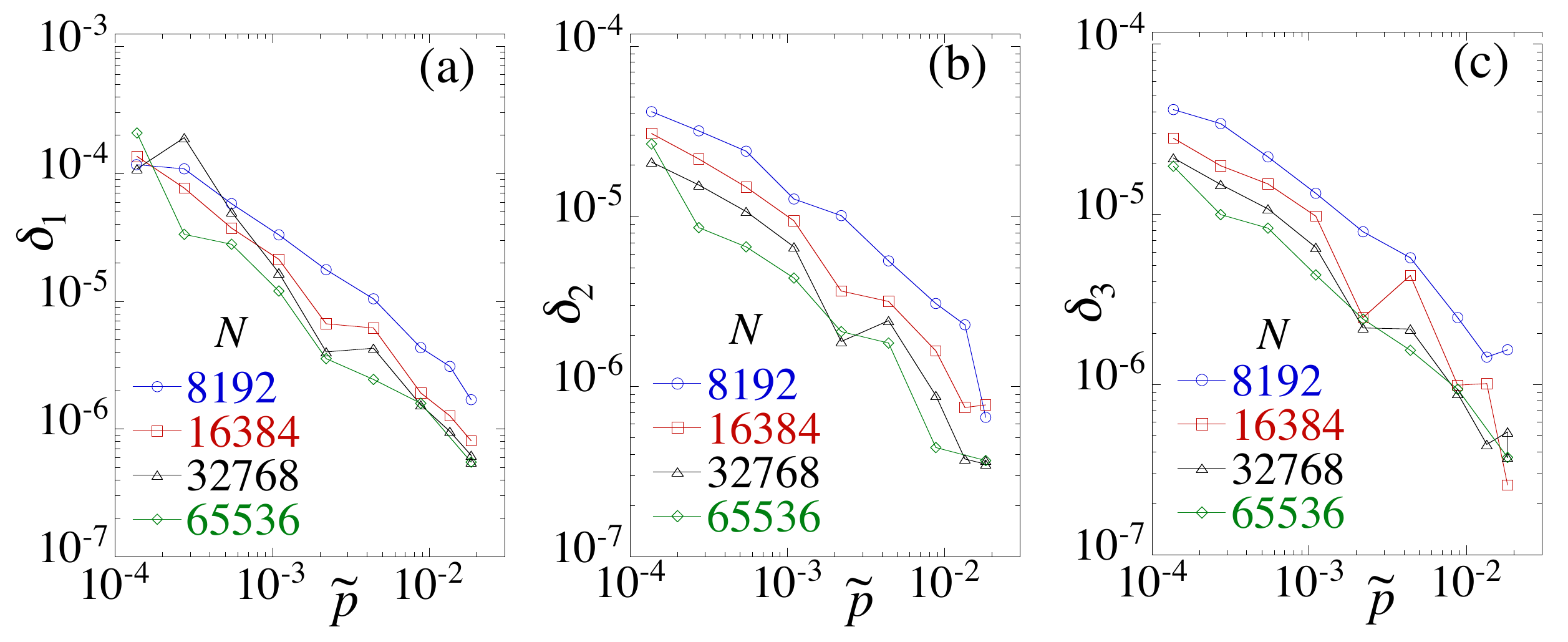}
\caption{(color online) Accuracy parameters (a) $\delta_1$, (b) $\delta_2$, and (c) $\delta_3$ of Eqs.~(\ref{delta1}), (\ref{delta2}) and (\ref{delta3}), that measure the relative deviations of the stress tensor $\Sigma_{\alpha\beta}$ from the target isotropic $\Gamma_N\delta_{\alpha\beta}$, vs stress per particle $\tilde p=\Gamma_N/N$, for system sizes $N=8102$ to $65536$.
}
\label{delta1-2-3}
\end{center}
\end{figure}

We first look at how well our procedure produces a packing with the desired isotropic global stress tensor, $\Sigma_{\alpha\beta}=\Gamma_N\delta_{\alpha\beta}$.  We compute the global stress tensor $\Sigma_{\alpha\beta}$ for our minimized configurations using the usual formula \cite{OHern} for a static frictionless system,
\begin{equation}
\Sigma_{\alpha\beta}=-\sum_{i<j}r_{ij\alpha}F_{ij\beta},
\end{equation}
where $\mathbf{r}_{ij}\equiv\mathbf{r}_j-\mathbf{r}_i$ is the center-to-center displacement from particle $i$ to particle $j$, $\mathbf{F}_{ij}=-\partial{\cal V}_{ij}/\partial\mathbf{r}_i$ is the contact force on $i$ due to $j$, and the sum is over all distinct pairs of particles in contact.
We then define three measures of the deviation of our minimized stress from the isotropic target value,
\begin{align}
\delta_1&\equiv \dfrac{\sqrt{\langle[\frac{1}{2}(\Sigma_{xx}+\Sigma_{yy})-\Gamma_N]^2\rangle}}{\Gamma_N}
\label{delta1}\\
\delta_2&\equiv\dfrac{\sqrt{\langle[\Sigma_{xx}-\Sigma_{yy}]^2\rangle}}{\Gamma_N}
\label{delta2}\\
\delta_3&\equiv\dfrac{\sqrt{\langle\Sigma_{xy}^2\rangle}}{\Gamma_N}.
\label{delta3}
\end{align}
$\delta_1$ measures the relative spread in the trace of $\Sigma_{\alpha\beta}$ about the target value $\Gamma_N$; $\delta_2$ measures the relative spread in anisotropy of the diagonal elements of $\Sigma_{\alpha\beta}$; and $\delta_3$ measures the relative spread in the off-diagonal elements of $\Sigma_{\alpha\beta}$.  In Fig.~\ref{delta1-2-3} we show our results for $\delta_1$, $\delta_2$ and $\delta_3$ vs $\tilde p=\Gamma_N/N$, for systems sizes $N=8192$ to $65536$.  We see that $\delta_1$ is less than $0.01\%$, while $\delta_2$ and $\delta_3$ are less than $0.004\%$, indicating a high accuracy in the desired stress tensor.  In all cases the accuracy improves as the stress per particle $\tilde p$ increases, and as the number of particles $N$ increases.

\begin{figure}[h]
\begin{center}
\includegraphics[width=3in]{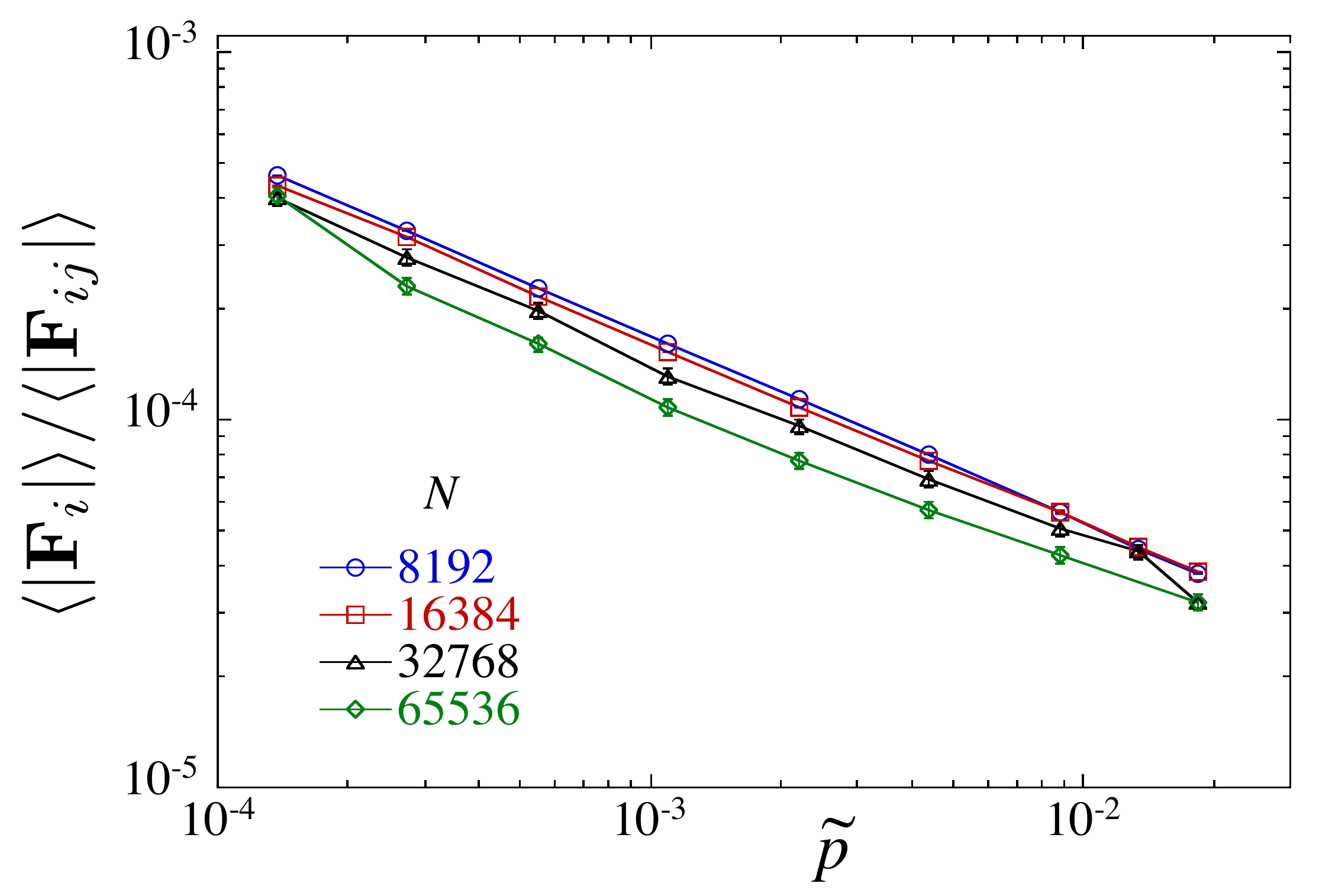}
\caption{(color online) Average magnitude of the residual net force on a particle $\mathbf{F}_i$, normalized by the average magnitude of the contact force $\mathbf{F}_{ij}$, at the termination of our minimization procedure.  Results are plotted vs the stress per particle $\tilde p=\Gamma_N/N$ for systems sizes $N=8192$ to $65536$.
}
\label{fi-o-fij-ave-vs-p}
\end{center}
\end{figure}

Next we look at how well our procedure produces a mechanically stable packing in which the net force on each particle vanishes.  The net force $\mathbf{F}_i$ on particle $i$ is just the sum over its contact forces, $\mathbf{F}_i=\sum_j \mathbf{F}_{ij}$.  In Fig.~\ref{fi-o-fij-ave-vs-p} we plot the average magnitude of the net force, normalized by the average magnitude of the contact force, $\langle|\mathbf{F}_i|\rangle/\langle|\mathbf{F}_{ij}|\rangle$, vs the stress per particle $\tilde p=\Gamma_N/N$, for system sizes $N=8102$ to $65536$.  We see that the residual net force on a particle at the end of our minimization procedure is less than $0.05\%$ of the average contact force.  This decreases as either $\tilde p$ or $N$ increases.

\begin{figure}[h]
\begin{center}
\includegraphics[width=3.4in]{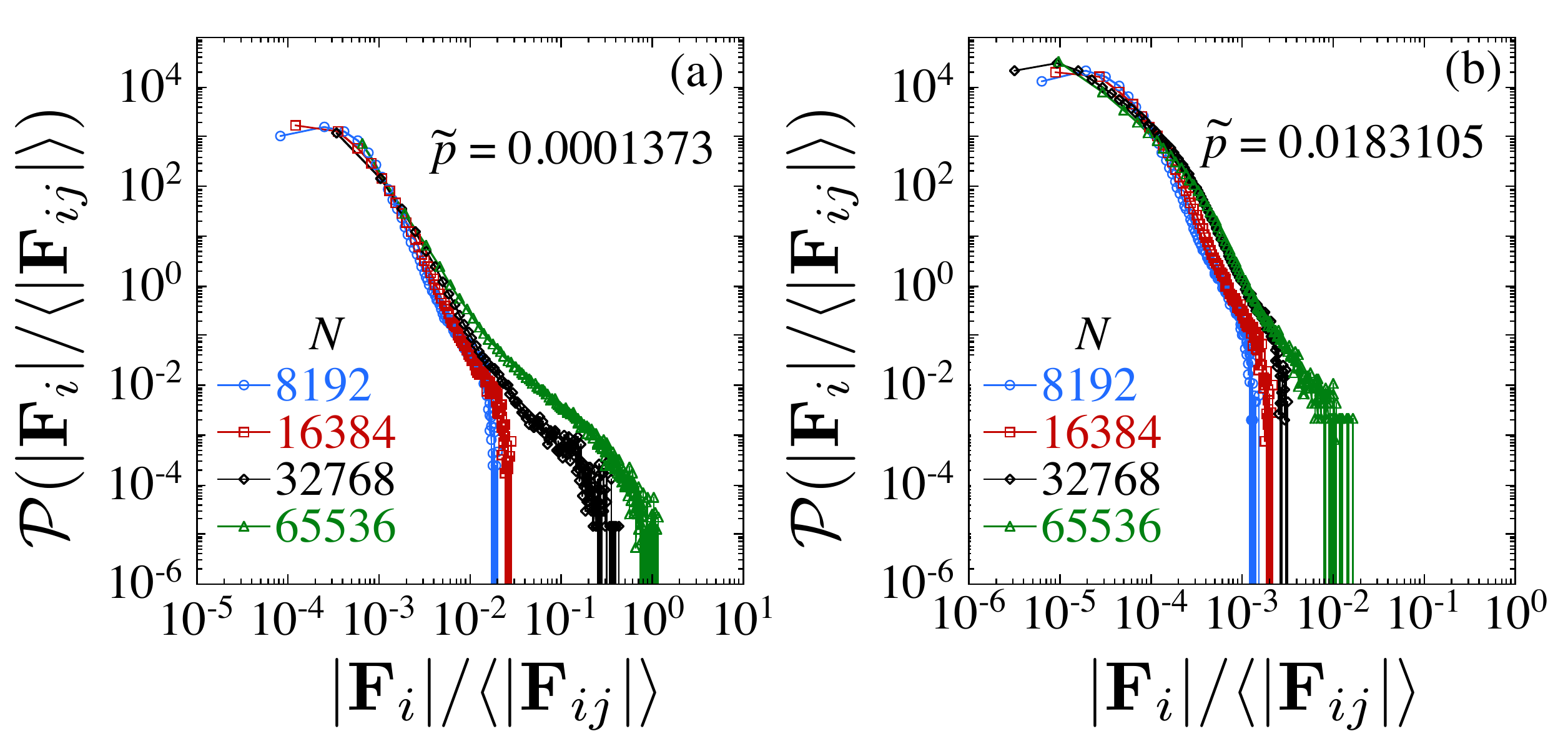}
\caption{(color online) Distribution ${\cal P}(|\mathbf{F}_i|/\langle|\mathbf{F}_{ij}|\rangle)$ of the residual net force on particles $|\mathbf{F}_i|$, normalized by the average magnitude of the contact force $|\mathbf{F}_{ij}|$, vs $|\mathbf{F}_i|/\langle|\mathbf{F}_{ij}|\rangle$, for different system sizes $N=8102$ to $65536$, at our (a) smallest $\tilde p=0.0001373$ and (b) largest $\tilde p=0.0183105$. 
}
\label{fi-o-fj-hist}
\end{center}
\end{figure}

Fig.~\ref{fi-o-fij-ave-vs-p} showed the {\em average} net residual force on particles.  In Fig.~\ref{fi-o-fj-hist} we show the distribution of such forces, ${\cal P}(|\mathbf{F}_i|/\langle|\mathbf{F}_{ij}|\rangle)$ vs $|\mathbf{F}_i|/\langle|\mathbf{F}_{ij}|\rangle$, for different system sizes $N=8102$ to $65536$, at our (a) smallest $\tilde p=0.0001373$ and (b) largest $\tilde p=0.0183105$.  We see that the large force tail grows as $N$ increases, but shrinks as $\tilde p$ increases.  For our largest system, $N=65536$, at our lowest stress per particle, $\tilde p=0.0001373$, there exist a very few particles whose net force is comparable to the average contact force.

\begin{figure}[h]
\begin{center}
\includegraphics[width=3.4in]{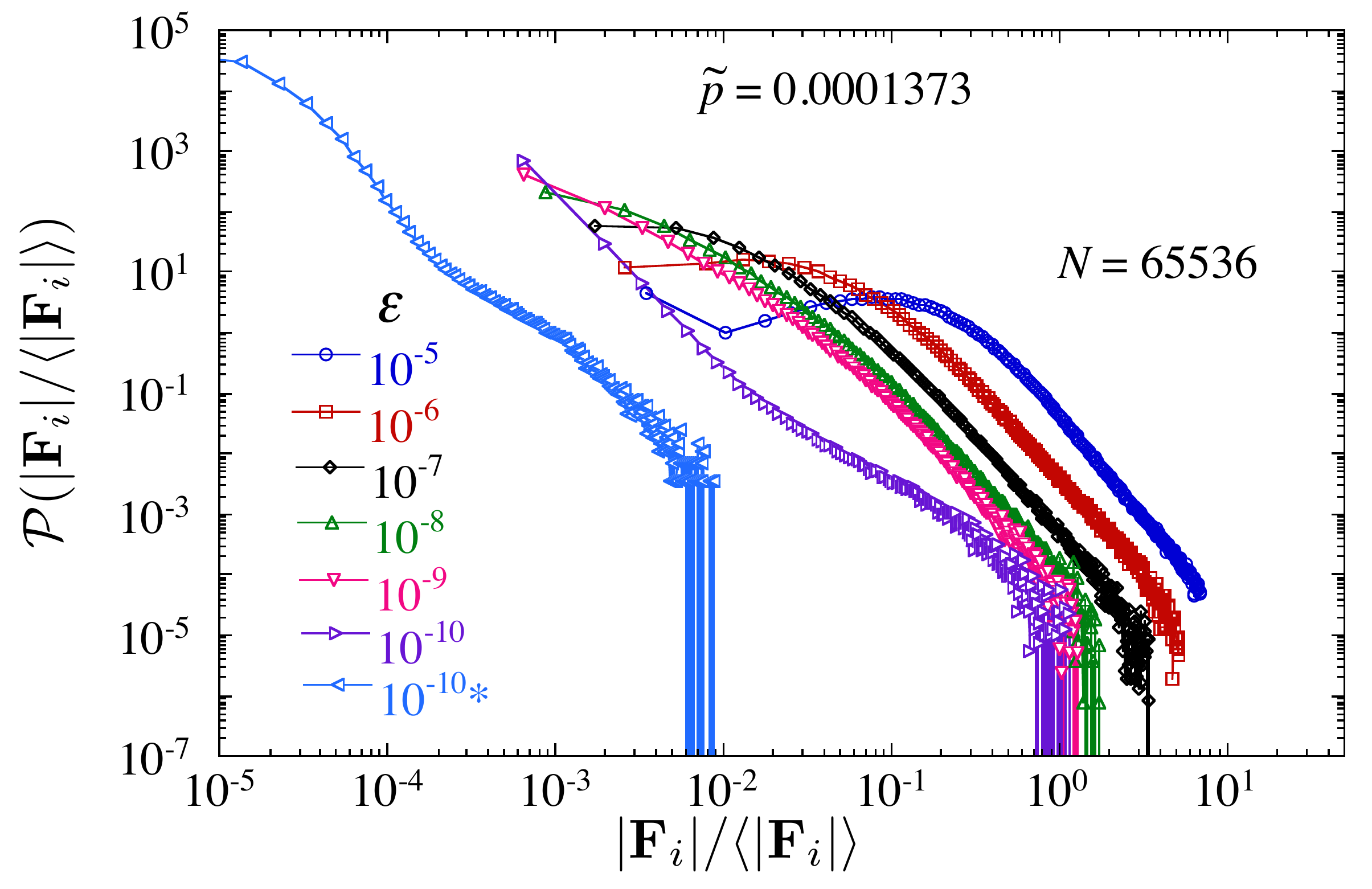}
\caption{(color online) Distribution ${\cal P}(|\mathbf{F}_i|/\langle|\mathbf{F}_{ij}|\rangle)$ of the residual net force on particles $|\mathbf{F}_i|$, normalized by the average magnitude of the contact force $|\mathbf{F}_{ij}|$, vs $|\mathbf{F}_i|/\langle|\mathbf{F}_{ij}|\rangle$, for different values of the accuracy parameter $\varepsilon$ that terminates our minimization of $\tilde U$.  Results are for a system of size $N=65536$ at $\tilde p=0.0001373$. Data labeled ``$10^{-10}*$" are results for a minimization of $\tilde U$ to accuracy $\varepsilon=10^{-10}$, followed by an additional minimization of interaction energy $U$ to accuracy $10^{-10}$ while holding the box parameters constant.
}
\label{fi-hist-eps}
\end{center}
\end{figure}

The average residual force $\langle |\mathbf{F}_i|\rangle$, and the large force tail of the distribution, is controlled by the accuracy parameter $\varepsilon$ that determines when we stop our minimization procedure, $(\tilde U_i-\tilde U_{i+50})/\tilde U_{i+50} <\varepsilon$.  In the body of this work, and in the above results, we have used $\varepsilon=10^{-10}$.  In Fig.~\ref{fi-hist-eps} we show the distribution ${\cal P}(|\mathbf{F}_i|/\langle|\mathbf{F}_{ij}|\rangle)$ for several different values of the accuracy parameter $10^{-10}\le\varepsilon\le 10^{-5}$, for our biggest system $N=65536$ at our lowest stress $\tilde p=0.0001373$.  We see that as $\varepsilon$ decreases, the average net force and the large force tail decrease.  Thus, as would be expected, decreasing $\varepsilon$ improves the accuracy of force balance on the particles in our minimized configurations.

We have attempted to improve upon the accuracy of force balance by adding a separate step of minimization in which, after the above criterion on $\tilde U$ is met, we then hold the box parameters $L_x$, $L_y$ and $\gamma$ constant while adjusting the particle positions to minimize the interaction energy, $(U_i-U_{i+50})/U_{i+50}<10^{-10}$.  The resulting distribution of net residual forces on particles is shown in Fig.~\ref{fi-hist-eps} labeled as ``$10^{-10}*$".  We find a significant reduction in the net force, with the average $\langle |\mathbf{F}_{i}|\rangle/\langle|\mathbf{F}_{ij}|\rangle$ decreasing roughly by a factor of 100.  However we also find that the accuracy of the system to have the desired target global stress decreases, with the parameters $\delta_i$ of Eqs.~(\ref{delta1}-\ref{delta3}) increasing roughly by a factor 10.  We have not tried to optimize the sequence of minimizing $\tilde U$ and $U$ as we have found our results for $\chi(\mathbf{q})$ to be insensitive to this additional step of minimization (see below), and so we have not used it for the results presented elsewhere in this paper.

\begin{figure}[h]
\begin{center}
\includegraphics[width=3.4in]{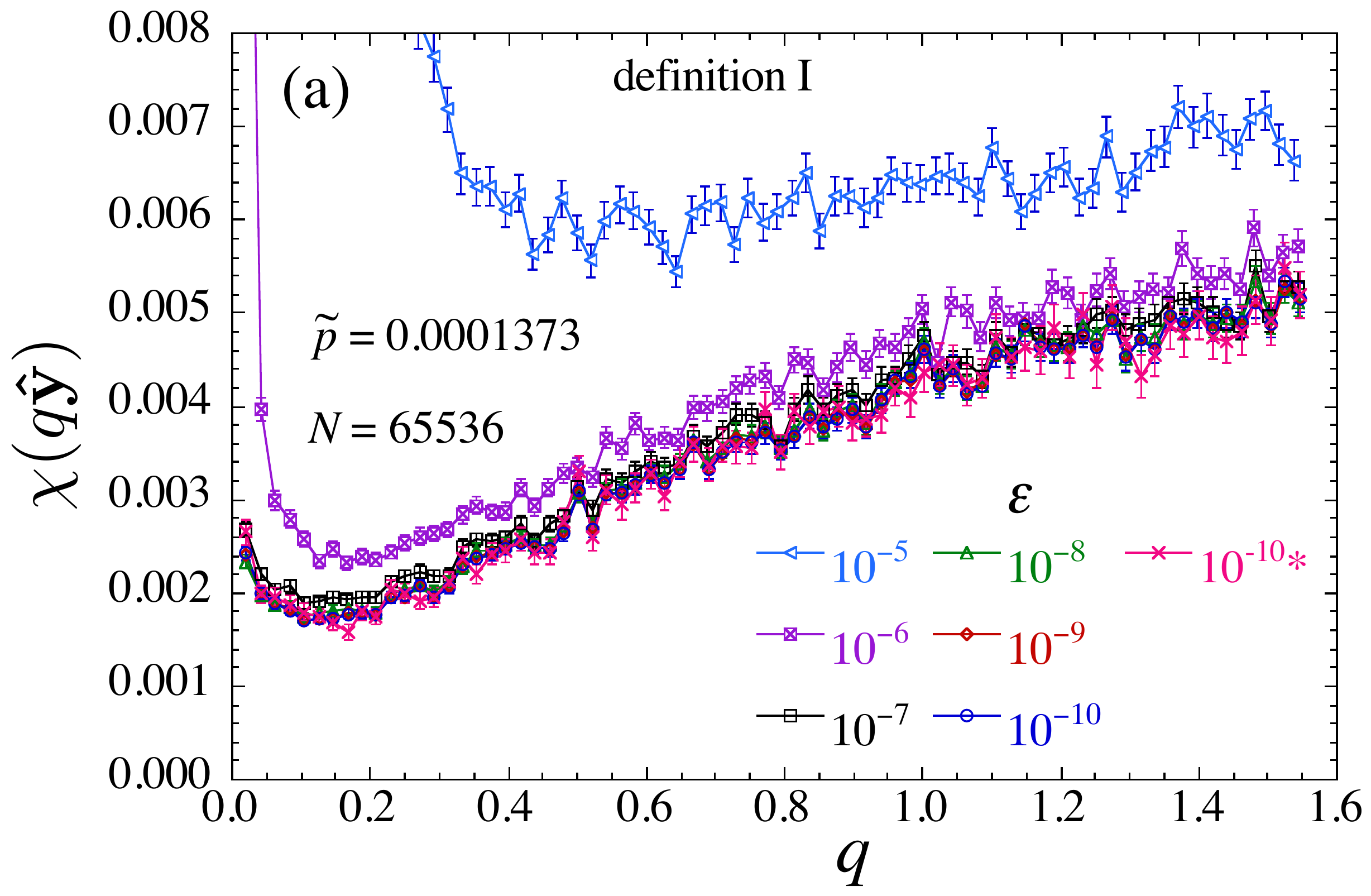}\\
\includegraphics[width=3.4in]{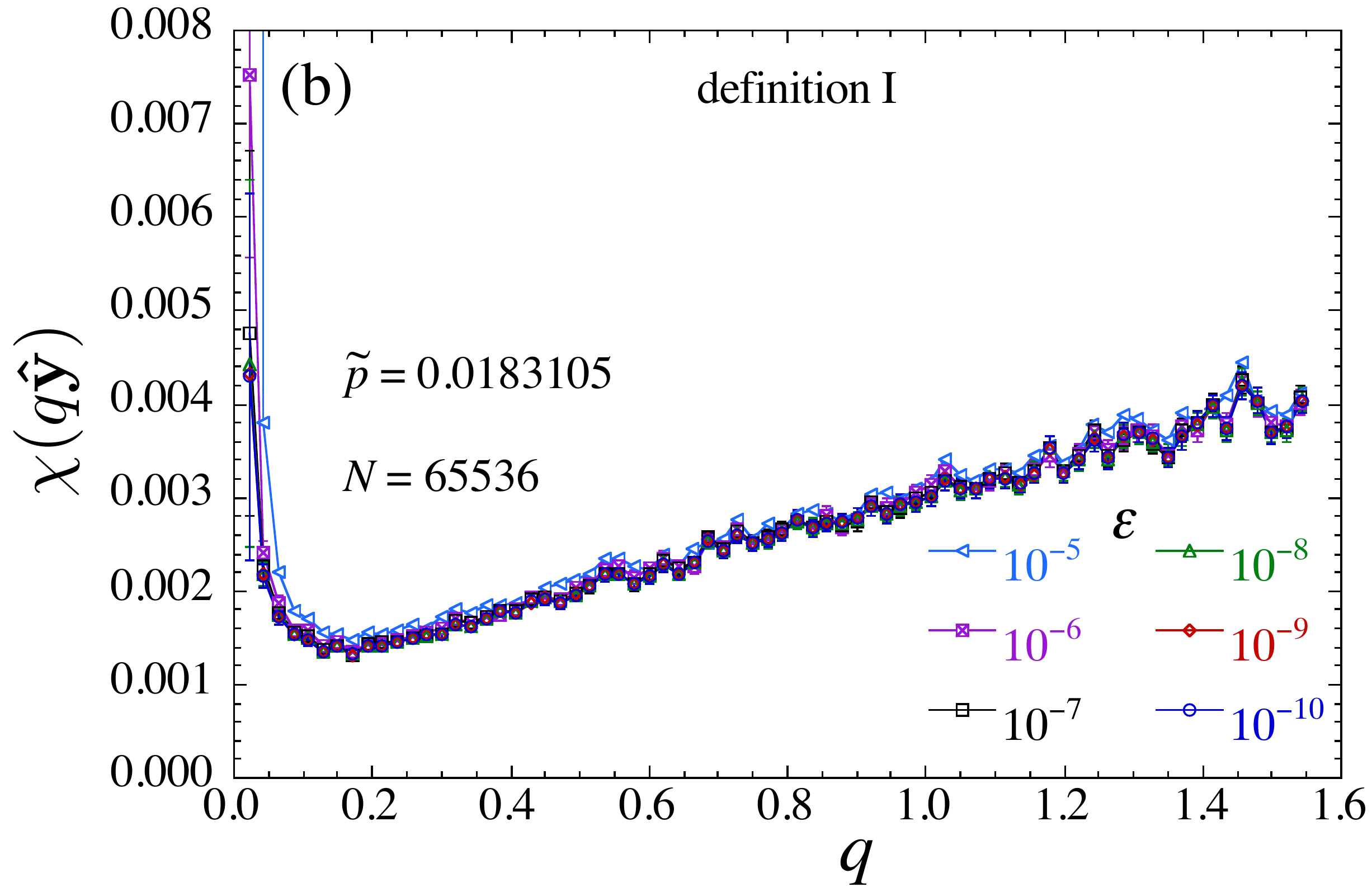}
\caption{(color online) Fluctuation in packing fraction $\chi(q\mathbf{\hat y})$ vs $q$, using definition I of Eq.~(\ref{edefI}) for the local packing fraction $\phi(\mathbf{r})$.  Results are shown for system size $N=65536$, for (a) our smallest value of the stress per particle, $\tilde p=0.0001373$, and (b) our largest value $\tilde p=0.0183105$.  We compare the values of $\chi(q\mathbf{\hat y})$ obtained when using different values of the parameter $\varepsilon$ that determines the stopping criterion for our minimization of $\tilde U$.  Data labeled ``$10^{-10}*$" in panel (a) are results for a minimization of $\tilde U$ to accuracy $\varepsilon=10^{-10}$, followed by an additional minimization of interaction energy $U$ to accuracy $10^{-10}$ while holding the box parameters constant.
}
\label{chiI-vs-q-eps}
\end{center}
\end{figure}

Finally, to determine whether the accuracy parameter $\varepsilon=10^{-10}$ used in this work is sufficient for our needs, we now check the sensitivity of $\chi(q\mathbf{\hat y})$ to the value of $\varepsilon$.  In Fig.~\ref{chiI-vs-q-eps} we plot $\chi(q\mathbf{\hat y})$ vs $q$ (using definition I of Eq.~(\ref{edefI}) for $\phi(\mathbf{r})$) for different values of $\varepsilon=10^{-5}$ to $10^{-10}$, for a system with $N=65536$ particles (we consider our largest system since that has the force distribution with the largest tail at large $|\mathbf{F}_i|$).  We show results for our smallest and largest values of the stress per particle $\tilde p=\Gamma_N/N$.  We see that if $\varepsilon$ is too large, the results at small $q$ are clearly dependent on $\varepsilon$.  But as $\varepsilon$ decreases, our results converge to a fixed $\varepsilon$-independent curve.  For the smallest  $\tilde p=0.0001373$ this happens for $\varepsilon\le 10^{-9}$, while for our largest $\tilde p=0.0183105$ we have convergence for $\varepsilon\le 10^{-8}$.  For the lowest $\tilde p$ in panel (a) we also show results for the case where $\varepsilon=10^{-10}$ and we add the second step of minimization described above, in which we fix the box parameters and only move particle positions to minimize $U$.  This data is labeled as ``$10^{-10}*$" in the figure.  We see that this additional step of minimization does not result in any noticeable change in $\chi(q\mathbf{\hat y})$.
We thus conclude that using $\varepsilon=10^{-10}$ with a single step minimization of $\tilde U$ gives sufficient accuracy for our needs.

\begin{thebibliography}{99}
\bibitem{note}Here we are referring to the definition of $\phi(\mathbf{q})$, and subsequent definition $I(q)$,  that is on page 4 of Ref.~[\onlinecite{Berthier}].
\bibitem{qdirection}For several cases we have explicitly computed $\chi(\mathbf{q})$ for $\mathbf{q}$ in other directions, specifically $\mathbf{q}=m\mathbf{b}_1$ and $\mathbf{q}=m(\mathbf{b}_1+\mathbf{b}_2)$.  We find that $\chi(\mathbf{q})$ depends only on the magnitude $|\mathbf{q}|$ for the small $|\mathbf{q}|\lesssim 1$ considered here.
\end{thebibliography}
\end{document}